\documentclass[a4paper,12pt]{article}
\usepackage{graphicx}
\usepackage{amssymb}
\usepackage{bm}
\usepackage{hyperref}
\usepackage{xcolor}
\usepackage{dsfont}
\usepackage{eurosym}
\usepackage{hyperref}

\usepackage{tikz}
\usepackage{amsmath}
\usepackage{stmaryrd}
\usepackage{caption}
\usepackage{subcaption}

\newcommand\blfootnote[1]{%
  \begingroup
  \renewcommand\thefootnote{}\footnote{#1}%
  \addtocounter{footnote}{-1}%
  \endgroup
}

\makeatletter
\let\@fnsymbol\@arabic
\makeatother

\begin{document}
\title{\LARGE A machine learning approach to support decision in insider trading  detection$^1$}

\author{Piero Mazzarisi$^{*,2,3}$, Adele Ravagnani$^{*,2}$, Paola Deriu$^4$,\\ Fabrizio Lillo$^{2,5}$, Francesca Medda$^{4,6}$, Antonio Russo$^{4}$}

\date{\today}

\maketitle
\abstract{Identifying market abuse activity from data on investors' trading activity is very challenging both for the data volume and for the low signal to noise ratio. Here we propose two complementary unsupervised machine learning methods to support market surveillance aimed at identifying potential insider trading activities. The first one uses clustering to identify, in the vicinity of a price sensitive event such as a takeover bid, discontinuities in the trading activity of an investor with respect to his/her own past trading history and on the present trading activity of his/her peers. The second unsupervised approach aims at identifying (small) groups of investors that act coherently around price sensitive events, pointing to potential insider rings, i.e. a group of synchronised traders displaying strong directional trading in rewarding position in a period before the price sensitive event. As a case study, we apply our methods to investor resolved data of Italian stocks around takeover bids.
\vspace{0.5cm}\\
\textbf{Keywords}: Insider trading, Market abuse, Unsupervised learning, Statistically validated networks}


\blfootnote{$^1$This paper represents the personal opinions of the authors and do not bind the membership organization in any way.\\
$^*$These authors contributed equally to the work.\\
$^2$Scuola Normale Superiore, Pisa\\
$^3$Dipartimento di Economia Politica e Statistica, Universit\`a di Siena, Italy.\\
$^4$Consob, Italy.\\
$^5$Dipartimento di Matematica, Universit\`a di Bologna, Italy.\\
$^6$University College London, Institute of Finance and Technology}

\newpage
\section{Introduction}

In financial markets the concept of market abuse consists in an intentional conduct that violates market integrity and natural demand-supply dynamics through misuse of privileged information, unlawful disclosure of inside information, unfair trading practices, price manipulation, creation of unfair market conditions, and deception of market players, to name but a few examples. 

In literature, the area of market abuse covers a number of different conducts that nonetheless could be grouped into two main categories: 1) insider dealing: the act of utilizing inside information in order to make, change, or cancel deals, or to encourage a third-party to deal using this knowledge and unlawful disclosure of inside information, by releasing information without correct permissions; 2) market manipulation, subdivided in trade base manipulation, action trade manipulation, or information based manipulation: in other terms an umbrella for a series of actions which distort market performance.

The European legislator defines \textit{market abuse} as any unlawful conduct on the financial markets involving the commission or attempted commission of insider dealing, i.e. the unlawful disclosure of inside information or the recommendation to others to engage in insider dealing, as well as market manipulation. Such conducts, preventing full and effective market integrity and compromising public confidence, are expressly prohibited and administratively sanctioned, leaving Member States the possibility of also imposing criminal sanctions.

In Europe the current legal framework of reference is represented by Directive 2014/57/EU (so-called MAD II), European Regulation 596/2014 (so-called MAR) as well as a bunch of Delegated Regulations supplementing the MAR Regulation with regard to regulatory technical standards on several aspects. The European legislator envisaged equipping the competent authorities of each Member State with adequate tools, powers and resources to ensure the effectiveness of supervision. In addition, the European provisions concerning market abuse require criminalisation of the most serious misconduct leaving to national legislators the power to criminalise certain misbehaviours. The Italian case is noteworthy because yet the implementation of the first Market Abuse Directive (MAD), in 2005,  was a hook to introduce severe administrative sanctions in addition to the pre-existing criminal penalties. Therefore, since long ago, in Italy Consob’s supervisory activities in this area may give rise to both administrative and criminal sanctioning proceedings, the latter by reporting the detected conducts to the Judicial Authority.

In this paper the focus is on insider trading, which is maybe the simplest market abuse conduct to conceive, but also one of the most widespread and difficult to enforce, since it is recognized as an illicit \textit{a probatio diabolica} for the difficulties inherent in the search for a smoking gun. Knowing in advance how the price will likely move in response to the release of the information to the market (a.k.a. price sensitive event (PSE), such as, for example, the announcement of a takeover bid), can be easily exploited to make a profit. Such a type of practice is prohibited or criminalized in most jurisdictions around the world \cite{bhattacharya2002}. However, rules are specific of each country and efforts in persecuting insider trading vary considerably.

The ``proof'' and the subsequent imposition of a sanction (either administrative or criminal) to a trader that have operated as an {\it insider} is however a complex process, involving many steps: (i) the detection of alerts pointing to anomalies that appear attributable to abusive behaviors , (ii) the concrete assessment of the allegedly suspicious conduct with respect to possible rationale that may have supported the strategy under analysis, (iii) the investigation phase aimed at gathering evidence and clues of the abusive conduct, and (iv) the subsequent legal trial to confirm the fact that the unlawful conduct was committed.

The scope of this work concerns the first step, i.e. the analysis of the trading behavior of investors in the presence of a price sensitive event, and, in part, the second one. The goal is providing a methodology based on unsupervised machine learning techniques that is able to provide and indication on whether the trading behavior of an investor or a group of investors is anomalous or not, thus supporting the monitoring and surveillance processes by the competent Authority and the assessment of the conduct.

In fact, the assessment of the trading behavior of an investor, alleged to be suspicious in terms of fault of abnormality remains a crucial point. In other words, focusing on a single trader, the smoking gun of the crime\footnote{As in a murder, finding the killer with the gun immediately after the fire shoot is a proof of guilt rarely available to investigators.} in the context of insider trading is a clue not easy to unearth; in addition also serious, precise and concordant evidences  could not be sufficient to build case strong enough. Nonetheless, once the alert is triggered, the more discontinuous an investor’s strategy is with respect to its trading history, the more suspicious its behavior can be considered. For example, the trading of only one stock {\it could} be suspicious when it happens: (i) for a specific time frame proximate to (and preceding) the dissemination of a price-sensitive news item, (ii) by assuming a rewarding position in relation to the price movement, and (iii) in the absence of other similar investments (i.e. in the same security over a longer timeframe or in securities similar in terms of capitalization, sector, risk, asset class, and for countervalues at least comparable in size to the investment under analysis).

Rather than splitting all investors into two classes, i.e. anomalous or not, it seems therefore more reasonable to assign a score to each trader as a measure of how much his/her behavior differs from the most suspicious one. Three considerations guided the design of our methodology:
\begin{itemize}
    \item First, anomalies need to be studied in a dynamical framework, in which any deviation from ordinary trading of an investor immediately before a price sensitive event must be characterized with respect to his/her activity in a past reference period when no anomalies are assumed. Such a comparison should include not only the trading in the asset under investigation but in whole market.
    \item Second, the deviations from ordinary trading of an investor should be compared with the trading behavior of his/her peers, defined as the investors which, in the past reference period, trade similarly to the investor under investigation. In other words, a modification in the trading behavior before a price sensitive event might require subtle interpretation  and assessment  if the (large) group of peers also modfiified their behavior in a similar way. For example, such cases might be due to some public leakage of information or some market dynamics which lead the investor and his/her peers to modify the trading.
    \item Third, when we look at small groups of traders, the coordination of a group in buying or selling a stock may be the signal of the spreading of confidential information within the group ({\it insider ring}). The study of such coordinated behaviors may further reveal when the confidential information started to spread before the price sensitive event. For instance, the appearance of coordinated buyers at some date before the announcement of a takeover bid could permit to infer from market data when the confidential information has been formed, then exploited by some insiders.
\end{itemize}

 The inference of such dynamic patterns can help the competent authority by providing a better understanding of the market dynamics and and can help in the identification of individual and collective suspects of insider trading activity.


{\bf Literature Review.} Market abuse detection naturally fits into the framework of anomaly detection, which basically amounts at identifying data instances that cannot be associated with a normal behavior and that are rare in the data set. The goal is to define a region of the features' space whose belong normal observations; observations that do not lie in this region are defined as anomalies \cite{AD_review}. Identifying this normality region is not straightforward: the boundary between normal and anomalous behavior is not always sharp, behaviors that are actually anomalous could be disguised in order not to be identified, the definition of normal behavior could be time varying and it is strongly dependent on the application domain, it is difficult to distinguish noise from anomalous behavior \cite{AD_review}. From a practical point of view, there are three main aspects which determine the formulation of the anomaly detection method: the nature of the input data, the type of anomaly, availability of data labels, the desired output of the technique. Data instances can be of various type (binary/categorical/continuous, univariate/multivariate) and independent among them or related to each other, as it is the case of time series and sequences, spatial data and graph data, for which ad hoc methodologies have to be employed \cite{AD_graph_review, aggarwal2013}. Concerning the type of anomalies, the standard case is represented by \textit{point anomalies}, which are single elements identified as anomalous; they could be \textit{global} or \textit{local} depending on whether the entire features' space or a specific region of it is considered \cite{AD_comparative_study}. Interestingly, there are cases when an element can be seen as normal, but when a given context is taken into account, it turns out to be an anomaly. We refer to this type as {\it contextual anomalies} \cite{AD_review}, also referred to as {\it conditional anomalies} \cite{song2007}. It may happen, for instance, that an investor has operated on a stock and, without a context, such an operation looks similar to other operations in the market. However, when compared to the own past behavior of the investor or to the operations of other investors, some discontinuity or synchronization patterns may be revealed and the operation could turn out to be identified as anomalous. Contextual anomalies problems can be tackled by algorithms for point anomaly detection once the context is included as a new feature. Finally, we could have data instances that are normal if considered individually, while they are anomalous together: they are the so-called \textit{collective anomalies} \cite{AD_review} and they can occur in data set where data instances are dependent. Another important issue is the availability of data labels which causes the employment of a different type of anomaly detection approach: supervised when each observation is labeled as normal or anomalous, semi-supervised when training data do not contain any anomalies and unsupervised when no labels are provided as in our interest case. Typically, outputs of anomaly detection algorithms associate to each observation a score, which quantifies the magnitude of its anomalous character. Setting a suited threshold, the ranked list of anomalies can provide labels for each data instance. 

It is evident anomaly detection problems are challenging and indeed, a variety of different approaches have been developed to tackle them. In particular, for the unsupervised approach, algorithms can be classified in four main categories: nearest-neighbor based, clustering-based, statistical and subspace techniques \cite{AD_review, AD_comparative_study}. Among them, the methods are multiple and their formulations are case-by-case dependent.

Anomaly detection has been widely explored in the literature, especially in the last years when its developments have been going at the same pace as machine learning's. The applications in the field of financial frauds detection are numerous \cite{west2016} and among them, some works are related to market abuse detection such as \cite{minenna2003, donoho2004, li2017, lamorgia2022}.  However, insider trading detection in stock markets is a quite unexplored topic.

{\bf Our contribution}\footnote{The methodology presented in the paper was conceived during 2022 for the purposes of developing a proof of concept. It is, in no way, a tool used in the analysis and investigations carried out by Consob. The methodology may possibly constitute in the future one of the tools to help and support the preliminary analysis and detection activities more efficiently. Any subsequent enforcement activity will, in any case, be based on the broader set of information that is gathered in the course of investigations and other possible types of analysis.}
In this paper, we propose a machine learning approach to the problem of insider trading detection, that leverages on the availability of a rich dataset including all daily transactions of all (anonymized) investors in the Italian stocks, from the beginning of 2019 to the third quarter of 2021. The methodology is based on two standard and well-known techniques, i.e. the k-means clustering algorithm \cite{kmeans} and the statistically validated co-occurrence networks \cite{svn1}, that are generalized in a dynamic framework for the study of contextual anomalies in the stock market. The anomaly detection approach proposed here is suited for the specific application to insider trading analysis The methodology has been built following the analysis and reasoning adopted by the Italian competent Authority (Consob) in the first steps of the investigation for insider trading, also taking into account the expertise  acquired by the analysis of the output cases already discussed in court.

In a first instance, we propose a dynamic clustering approach based on the {\it k-means} algorithm \cite{kmeans}. The main idea is representing market data in an Euclidean space in which the trading activity on one stock by an investor in a given time period is described by a point whose coordinates are some trading  features, namely the total bought or sold {\it turnover}, the trading concentration (also called {\it magnitude}), and the financial {\it exposure} on the stock. The clustering of investors in such feature space at successive times within a given reference period allows to achieve a stable description of the usual trading behavior, described by clusters persistent in time. Such a characterization of the market provides the ideal context in which we can assess the discontinuity of the investors' behavior, if any, {\it both with respect to his/her past trading behavior and with respect to the trading behavior before the PSE of his/her peers}. For example, when a price sensitive event for the stock is within the reference period, some operations that places an investor in a rewarding position never assumed before, i.e. the switching towards the cluster he/she never belonged to, can be classified as anomalous. The goal is then revealing all discontinuities towards rewarding position with respect to the price sensitive event and ranking them according to some anomaly score, thus providing the competent authority with a list of suspicious behaviors to be further analyzed.

The second clustering approach we employ is based on the so-called Statistically Validated Networks (hereinafter SVN). This is an unsupervised learning method which was introduced in \cite{svn1} and further employed in other works such as \cite{svn2}. This method aims at detecting unexpected structures in the projection of a bipartite network which represent a complex system. Similarly to \cite{svn2}, we start from a bipartite network investors-trading days, which represents the trading activity for a specific asset in the Italian Stock Exchange. This network is projected in a traders' graph where links are statistically validated, thus capturing agents' trading co-occurrences. Then, on this validated projected network, clusters are identified and they represent groups of traders who are synchronized in the kind and time of trading actions.  Given these clusters, our analysis focuses on a time window before the PSE under investigation and our goal is to detect group of traders who are likely to be suspicious for market abuse. If for instance, we are studying a takeover bid, investors who belong to a group with suspicious behavior, trade in a rewarding manner in the reference period before the price sensitive event. Since a positive reaction from the market follows a takeover bid, their trading is extremely polarized towards buying transactions.

As it is illustrated in the following, the SVN-based clustering approach turns out to be a powerful tool to detect group of investors with suspicious trading before a PSE. Moreover, capturing synchronized trading activity could be of help for the competent authority in establishing when information becomes spread among the market operators and so, how the reference period to focus on market abuse investigations should be framed. 

It is worth noticing that given the different nature and goals of the k-means and SVN clustering approaches, these two methods lead to different and complementary results, potentially revealing the presence of single insiders and/or insider rings. Finally, the coupled use of the two methods needs to be intended as a support tool to the competent authority in the first steps of insider trading analysis. As such, the proposed methodology contributes to the field of {\it human-centered} decision support systems \cite{ejor_barthelemy}. In fact, the SVN-based approach can also be used in a \textit{human-in-the-loop} manner \cite{wu}: a trader $z$ who is considered as suspicious can be used as seed in the statistically validated network of investors to find other possible suspects (neighbors of $z$ in the SVN displaying coordinated trading behavior) or even potential insider rings (clique of connected nodes in the SVN displaying high synchronization).


\bigskip
{\bf Paper organization.} The paper is organized as it follows. Section \ref{sec:methods} describes the two proposed methods for the identification of individual and collective potential insider trading activities. Section \ref{sec:data} presents the dataset we use in our empirical analysis and Section \ref{sec:results} presents the results obtained with our method, with a special focus on one PSE. Results for other PSEs are presented in the Appendix. Finally, Section \ref{sec:conclusions} draw some conclusions and present some suggestions for further work.

\section{Methods}\label{sec:methods}
In this Section, we present two different clustering methods, which allow to find groups of investors with similar behavior in trading a given stock, and generalize them to devise a methodology of contextual anomaly detection for insider trading. In the first method, groups are identified by partitioning investors depending on some trading features (turnover, magnitudo, and exposure). In the second method, investors are identified as similar when they trade in a coordinated way (e.g. they buy or sell in the same days), displaying some significant correlation in the trading activity. Finally, contextual anomalies refer to either the discontinuity and the coordination of the trading behavior of investors in the presence of a price sensitive event. This is shown for a particular case study in the next section.

\subsection{Method based on k-means clustering}\label{sec_kmeans} The {\it k-means} clustering algorithm \cite{kmeans} is an unsupervised learning method for finding clusters of points in a set of unlabeled data that lie in a Euclidean space. Each data point $x$ is a {\it $n$-tuple} of real numbers characterizing the trading features of each investor. The method aims to partition $N$ data points into $K$ clusters, each point belonging to the cluster with the nearest mean, which is named the {\it centroid} of the cluster. Such a particular point in the feature space summarizes the average characteristics of all points in the cluster. The output of the clustering algorithm is a partition of the feature space into {\it Voronoi cells}. More specifically, given a set of data points $\{x_i\}_{i=1,\ldots,N}$ with $x_i\in\mathbb{R}^n$ and the number $K$ of clusters in which we aim to partition the feature space, the $k-means$ algorithm finds $K$ sets $\mathcal{S}=\{S_k\}_{k=1,\ldots,K}$ of points corresponding to $K$ centroids, in such a way the squared distance (variance) of points from the centroid within each cluster is minimized, i.e.
\begin{equation}\label{kmeans}
\mbox{argmin}_{\mathcal{S}}\sum_{k=1}^K\sum_{i\in S_k}\vert\vert x_i - c_k\vert\vert^2 = \mbox{argmin}_{\mathcal{S}}\sum_{k=1}^K\vert S_k\vert \mbox{Var}(S_k)
\end{equation}
where $c_k=\frac{1}{\vert S_k\vert}\sum_{i\in S_k}x_i$ is the mean of points belonging to $S_k$, whose cardinality is $\vert S_k\vert$. The vector $c_k$ defines the position of the centroid in the feature space.

The problem (\ref{kmeans}) is in general computationally difficult (NP-hard), however there exist several heuristic methods converging quickly to a local minimum. Here, we use a two-phase iterative algorithm, see e.g. \cite{lloyd}, that minimizes the within-cluster variances by alternating {\it batch updates} (reassigning points to their nearest cluster centroid, all at once, followed by recalculation of cluster centroids) and {\it online updates} (reassigning points one by one only if it reduces the within-cluster variances).

\subsubsection{Trading features} Let $N$ be the number of investors, $M$ the number of stocks, and $S$ the number of trading days. For a given stock $j\in\{1,\ldots,M\}$ and a given time window $\Delta = \left(t,t+\vert\Delta\vert\right]$ with $\vert\Delta\vert>0$, e.g. a week ($\vert\Delta\vert=5$) or a month ($\vert\Delta\vert=20$), each investor $i$ is associated with a triple of features (i.e. $K=3$), summarizing his/her trading activity during that period:
\begin{enumerate}
    \item the {\it signed turnover}, namely the aggregated Euro turnover of operations within the period, with positive (negative) sign for a net buying (selling) volume,
    $$
    A_i^{(j)} = \sum_{t\in\Delta} A_b(i,j,t) - \sum_{t\in\Delta} A_s(i,j,t),
    $$
    where $A_b(i,j,t)$ and $A_s(i,j,t)$ are the total Euro turnover bought and sold by investor $i$ for stock $j$ in day $t$;
    \item the {\it magnitudo} (or portfolio concentration), namely the relative Euro turnover (without sign) traded in the stock with respect to the total amount traded in any stock,
    $$
    a_i^{(j)} = \frac{\tilde{A}_i^{(j)}}{\sum_{j=1}^M\tilde{A}_i^{(j)}}, \:\:\:\mbox{with}\:\:\:\tilde{A}_i^{(j)} = \sum_{t\in\Delta} A_b(i,j,t) + \sum_{t\in\Delta} A_s(i,j,t);
    $$
    \item the {\it maximum exposure} (in Euro) to the stock in the time window,
    $$
    E_i^{(j)} = \left(\max_{t\in\Delta}\vert\alpha_t\vert\right)\mbox{sign}(\alpha_{\tilde{t}})
    $$
where $\alpha_t$ is the position\footnote{In general, information on the precise composition of the portfolio of each investor is not available, but only the daily aggregated information on his/her operations. As a proxy of asset positions we assume that they are zero on Jan. 1, 2019, then each position at day $t$ is obtained by aggregating bought and sold turnovers from Jan. 1, 2019 up to $t$.} (in Euro turnover) of investor $i$ in stock $j$ at day $t$ and $\tilde{t} = \mbox{argmax}_{t\in\Delta} \vert\alpha_t\vert$.
\end{enumerate}

For a stock $j$, a data point having the three features as coordinates in an Euclidean space $\mathbb{R}^3$ describes the trading behavior of an investor $i$, aggregated over a time window $\Delta$. 

While the magnitudo (i.e. relative turnover) $a_i^{(j)}$ is  by definition between $0$ and $1$, both the aggregated turnover and the exposure depend strongly on the size of the trader's portfolio. In absence of any normalization, the clustering algorithm would tend to group together investors of similar sizes, independently from their idiosyncratic trading behavior. Hence, in order to avoid such a spurious effect, it is convenient to re-scale each value within the interval $\left[-1,1\right]$, as follows. First, let us consider the whole period $[t_1,t_S]$, containing $m>1$ time windows of length $\vert\Delta\vert$, possibly overlapping.\footnote{An example is a reference period of one year, with twenty days long windows (i.e. one business month), which are rolled week by week starting from January.} Then we compute the three features for each investor $i$ within each time window, i.e. $x_i\equiv \{A_{i,s}^{(j)}, a_{i,s}^{(j)},E_{i,s}^{(j)}\}_{s=1,\ldots,m}$. Finally, we define the re-scaled signed turnover and the re-scaled maximum exposure as 
$$
A_{i,s}^{(j)} \mapsfrom \frac{A_{i,s}^{(j)}}{\max_s\vert A_{i,s}^{(j)}\vert},
$$
and 
$$
E_{i,s}^{(j)}\mapsfrom \frac{E_{i,s}^{(j)}}{\max_s\vert E_{i,s}^{(j)}\vert}.
$$

\begin{figure}
    \centering
    \includegraphics[width=.55\linewidth]{ 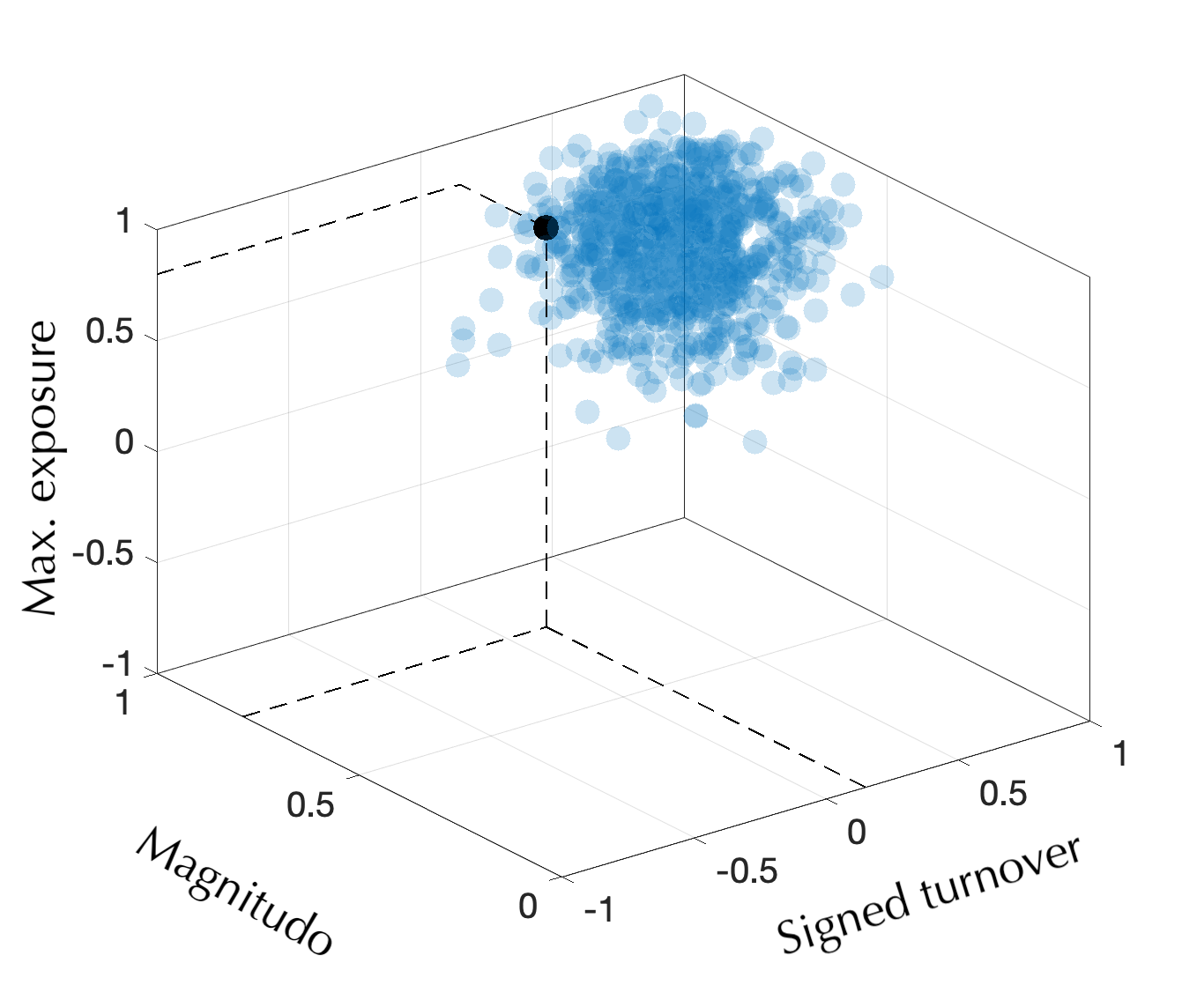}
    \caption{A pictorial representation of the trading activity (black dot) of an investor by using the three features, i.e. signed turnover, magnitudo, and maximum exposure, as defined in the main text.}
    \label{example_kmeans}
\end{figure}

In this way, we are able to map the trading activity of an investor $i$ on a stock $j$ in a given time window to three normalized features defined in the space $[-1,1]\times [0,1]\times [-1,-1]$. A pictorial example is shown in Figure \ref{example_kmeans}, where a generic investor, i.e. the black dot, belonging to some cluster is identified by three coordinates in the feature space, namely the signed turnover, the magnitudo, and the maximum exposure as defined above.

Such an approach is particularly useful when we aim to characterize the trading behavior of an investor within a given time window, with respect to the typical operations done in the whole reference period. In particular, when $t_S$ is the date of a price sensitive event for a stock, it is possible to compare the trading behavior of an investor in the previous $\vert\Delta\vert$ days with the past. From the point of view of the regulator, this serves to verify whether the current behavior of an investor is coherent or not with its own past.


Finally, notice that
the clustering analysis for the time window $\Delta$ includes only those investors that are active at least one day in $\Delta$. 

\subsubsection{Number of clusters} The number of clusters $K$ in which we partition the feature space is the only input we give to the clustering algorithm. In absence of any prior information on the number of clusters, this represents an hyperparameter one needs to optimize. This is usually done by following some heuristics \cite{tibi}. Here, we use a standard approach in clustering analysis known as the {\it elbow} method. It is based on the computation of the mean variance of features $\{x_i\}_{i=1,\ldots,N}$ with respect to the centroids' position $\{c_k\}_{k=1,\ldots,K}$ as a function of the number of clusters $K$, namely
\begin{equation}\label{elbow}
\mathcal{E}_K = \frac{1}{K}\sum_{k=1}^K\sum_{i\in S_k}\vert\vert x_i-c_k\vert\vert^2.
\end{equation}
In fact, $\mathcal{E}_K$ is a non-increasing function in $K$, since a larger number of clusters will naturally improve the fit, thus reducing the variance. However, this leads to over-fitting at some point. The idea is finding a cutoff point at which the diminishing of the variance (\ref{elbow}) becomes negligible. This is obtained by finding the number of clusters $K$ such that the relative (negative) increments of $\mathcal{E}_k$ are smaller than $5\%$ 
when more than $K$ clusters are used. The procedure is in general repeated within each time window of length $\vert\Delta\vert$, by covering the whole period $[1,T]$. We select the optimal number of clusters $K$ as the (rounded) mean value over the $m$ time windows.

\subsubsection{Dynamic clustering}
The {\it k-means} clustering algorithm is applied to the set of features computed within each time period $\Delta$, by rolling the window over the whole period $[1,T]$. For each window, the output of the method is the positions of centroids, together with the vector of labels, indicating which investors belong to each cluster. Unfortunately, the convergence to a global minimum is not ensured, possibly resulting in very different centroids' positions moving from one time period to the next one. Moreover, any permutation of labels does not change the loss function in Eq. (\ref{kmeans}). However, when we roll the time window over the whole period, it is convenient to associate two clusters in a row  with the same label when both of them are formed in large part by the same investors. 

In order to recover a pattern of stability for clusters, the centroids' positions that solve the minimization problem at a given time are then used as starting point of the clustering method when the time window is shifted one step forward\footnote{At initial time, random positions are used. The algorithm is then repeated many times for different input seeds, in order to find the global minimum over the number of initializations.}. Once the new centroids' positions are found in the new time window, we consider all possible permutations of labels over the set $\{1,\ldots,K\}$ and compute the Jaccard similarity coefficient \cite{tibi} between the current clusters and the previous ones. This is a metric measuring the overlap between the two sets of elements, which is equal to one when each element is also in the other set and vice versa, zero if not. Any other value between zero and one suggests a partial overlapping between the two clusters, such that the larger is the overlapping, the larger is the coefficient. The final assigned labels are the ones maximizing the Jaccard similarity. In this way, cluster stability tends to be preserved over time.

\subsubsection{Identification and classification of potential insiders.}
Once the dynamic clusters have been obtained, the method identifies the potential insider investors. To a PSE one can associate rewarding position (buy or sell) which tells us which is the trading direction that would have produced a profit. Let us assume that in the PSE under investigation the rewarding position is buy. An insider $i$ that aims at maximizing the profit by exploiting the sensitive information on the takeover bid would purchase during the reference period the largest possible volume ($A_i\rightarrow 1$), in particular concentrating his/her investment on the stock ($a_i\rightarrow 1$), thus reaching his/her maximal historical exposure ($E_i\rightarrow 1$). It is clear that the best rewarding position in the feature space is represented by the unit vector $\bm{1}\equiv(1,1,1)$.\footnote{Notice here that such considerations are independent from the size of the investor. Two investors with different capital, but trading similarly in a given period, are considered similar in the feature space. This is coherent with the current regulation about insider trading, which persecutes the illegal behavior, independently from the volume of the investment or the profit, if any.} This allows us to identify the cluster with rewarding position (w.r.t. PSE). 

Following the principles exposed in the Introduction, an investor that is suspect for insider trading has a discontinuous behavior with respect to his/her own past trading and with respect to the trading behavior of her peers. The proposed clustering method permits to identify the set of suspects with discontinuous trading behavior in a data-driven manner. In fact, an investor is defined as discontinuous if he/she has never belonged to the cluster in the past (non-overlapping) time windows.\footnote{Before such an analysis, we need to ensure the stability in time of the cluster under investigation. This is done by verifying that the position of the centroid is almost constant for all time windows.} If this is the case, there are two mutually exclusive possibilities: (i) the investor has never traded the stock in the considered period; (ii) the investor has traded the stock in the past, but taking a different position, e.g. lower exposure ($E_i\ll 1$) and larger portfolio diversification ($a_i\ll 1$). We name the first type of investors as {\it hard discontinuous} traders, while the second one as {\it soft discontinuous}.

Finally, given the identified set of discontinuous traders, it is of interest to obtain a ranking of the suspects depending on some score function measuring how much anomalous is their trading behavior. To this end, we can consider any decreasing function of the Euclidean distance $d(x_i,\bm{1})\equiv\sqrt{\vert\vert x_i-\bm{1} \vert\vert^2}$ of the vector of features $x_i$ characterizing the investor $i$ from the best rewarding position $\bm{1}$. In the empirical analyses below, the score is defined as $s_i = \exp(-d(x_i,\bm{1}))$, but other functions can be used.



\subsection{Method based on Statistically validated Networks}\label{method_svn}
The second clustering approach we employ is based on the so-called Statistically Validated Networks, which is an unsupervised learning method introduced in \cite{svn1} and further employed in other works such as \cite{svn2,baltakiene,musciotto}. 

This method aims at detecting structures in the projection of bipartite networks which represent complex systems. Analogously to \cite{svn2}, the complex system under our investigation is the activity of traders in the Italian Stock Exchange. Our goal is to identify clusters of investors who are synchronized in the kind and time of trading actions. This clustering is the starting point for a subsequent analysis which allows us to detect group of traders who are likely to be suspicious for market abuse in correspondence of price sensitive events.

\subsubsection{The SVN}\label{subsection_method}
\begin{figure}
\centering
\begin{tikzpicture}[roundnode/.style={circle, draw=gray!60, fill= gray!5, very thick, minimum size = 1cm},
squarednode/.style= {draw = white, rectangle, minimum size = 2cm}]
\node[roundnode] (1) at (2,4) {ID 1};
\node[roundnode] (2) at (4,4) {ID 2};
\node[roundnode] (3) at (6,4) {ID 3};
\node[roundnode] (4) at (8,4) {ID 4};
\node[roundnode] (1) at (1,0) {1};
\node[roundnode] (2) at (2,0) {2};
\node[roundnode] (3) at (3,0) {3};
\node[roundnode] (4) at (4,0) {4};
\node[roundnode] (5) at (5,0) {5};
\node[roundnode] (5) at (6,0) {6};
\node[roundnode] (5) at (7,0) {7};
\node[roundnode] (5) at (8,0) {8};
\node[roundnode] (5) at (9,0) {9};
\node[roundnode] (5) at (10,0) {10};
\node[roundnode] (5) at (11,0) {11};
\node[roundnode] (5) at (12,0) {12};
\draw[red] (2,3.5) -- (1,0.5);
\draw[red] (2,3.5) -- (2,0.5);
\draw[red] (2,3.5) -- (3,0.5);
\draw[red] (2,3.5) -- (4,0.5);
\draw[red] (2,3.5) -- (5,0.5);

\draw[red] (8,3.5) -- (1,0.5);
\draw[red] (8,3.5) -- (2,0.5);
\draw[red] (8,3.5) -- (3,0.5);
\draw[red] (8,3.5) -- (4,0.5);
\draw[red] (8,3.5) -- (5,0.5);

\draw[green] (6,3.5) -- (8,0.5);
\draw[green] (6,3.5) -- (9,0.5);
\draw[green] (6,3.5) -- (10,0.5);
\draw[green] (6,3.5) -- (11,0.5);
\draw[green] (6,3.5) -- (12,0.5);

\draw[green] (4,3.5) -- (8,0.5);
\draw[green] (4,3.5) -- (9,0.5);
\draw[green] (4,3.5) -- (10,0.5);
\draw[green] (4,3.5) -- (11,0.5);
\draw[green] (4,3.5) -- (12,0.5);

\draw[blue] (2,3.5) -- (11,0.5);
\draw[blue] (2,3.5) -- (10,0.5);
\draw[green] (2,3.5) -- (7,0.5);

\draw[blue] (4,3.5) -- (1,0.5);
\draw[green] (4,3.5) -- (2,0.5);
\draw[green] (4,3.5) -- (3,0.5);
\draw[red] (4,3.5) -- (5,0.5);
\draw[red] (4,3.5) -- (6,0.5);

\draw[green] (6,3.5) -- (3,0.5);
\draw[green] (6,3.5) -- (2,0.5);

\draw[blue] (8,3.5) -- (12,0.5);
\draw[green] (8,3.5) -- (11,0.5);
\draw[blue] (8,3.5) -- (10,0.5);

\node[squarednode] (11) at (14,0) {Trading days};
\node[squarednode] (12) at (14,4) {Investors};
\end{tikzpicture}
\caption{Example of a bipartite network. The first layer of nodes (on the top) is made up of $4$ investors and the second layer (on the bottom) of $12$ trading days. Three types of edge are allowed: red = \textit{buying} state, green = \textit{selling} state, blue = \textit{buying-selling} state.}
\label{example_bipartite}
\end{figure}
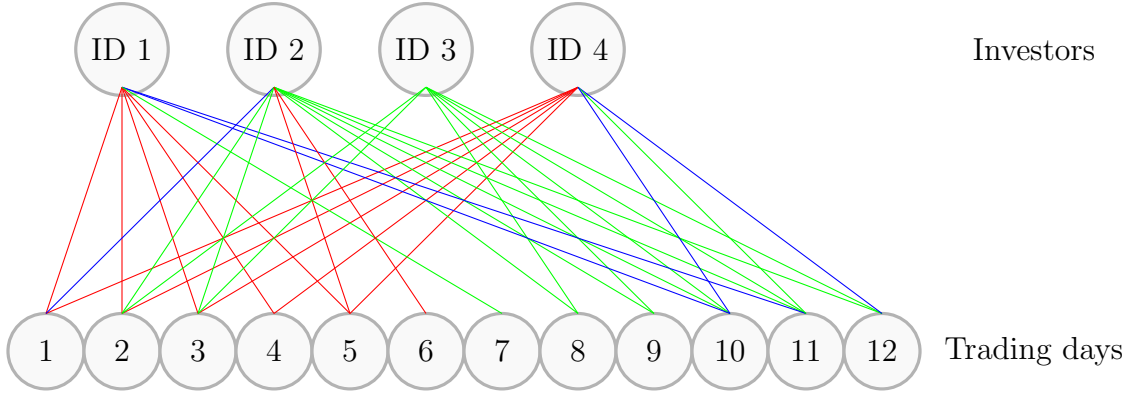

Given the stock under investigation, the first step of the method consists in constructing a bipartite network, where the two sets of nodes are made up of traders and trading days respectively.  Only links between a trader and a trading day can occur and they represent trading activity states. Figure \ref{example_bipartite} provides an example of this bipartite network. Each edge colors stands for a different trading state. 

More formally, let us consider $i=1\ldots, N$ traders and $t=1,\ldots,T$ trading days. In order to characterize investors' trading activity, the following metric, which we denominate \textit{directionality}, is associated to investor $i$ in day $t$:
    \begin{equation}\label{eqr}
        r(i,t) = \frac{V_b(i,t) - V_s(i,t)}{V_b(i,t) + V_s(i,t)}
    \end{equation}
where $V_b(i,t)$ and $V_s(i,t)$ are the total volumes (in shares) bought and sold by trader $i$ in day $t$. The metric $r(i,t)$ is normalized in the sense that it can assume values between $-1$ and $1$. This overcomes the difficulty of comparing heterogeneous investors who trade volumes with different orders of magnitude and the activity of a single investor which can be heterogeneous over time.

Depending on the value of this variable $r(i,t)$ compared to a fixed threshold $\theta$, three different states are defined:
    \begin{itemize}
        \item $r(i,t) > \theta$, \textit{buying} state i.e. $s(i,t) = b$;
        \item $r(i,t) < -\theta$, \textit{selling} state i.e. $s(i,t) = s$;
        \item $-\theta \leq r(i,t) \leq \theta$, \textit{buying-selling} state i.e. $s(i,t) = bs$.
    \end{itemize}

Concerning the threshold $\theta$, in the following we set $\theta = 0.01$ but, as also observed in \cite{svn2}, we verified that results are not affected by the choice of this parameter.

Given the states $\{s(i,t) \ i=1,\ldots,N, \ t=1,\ldots,T \}$, we build a bipartite network where 
\begin{itemize}
    \item one layer is made up of traders: $A = \{1,\ldots,N\}$;
    \item the other layer is made up of trading days: $B = \{1,\ldots,T\}$;
    \item only links of the type $(i,t)$ $i \in A$, $t \in B$ are admitted;
    \item each link can be $b$, $s$, $bs$, depending on $s(i,t)$.
\end{itemize}

The goal of our analysis is to perform a clustering of traders therefore, starting from our bipartite system, we focus on the projected network of traders, which is a new network with $N$ nodes, each representing a trader. A link between node $i$ and node $j$ exists if in the bipartite network, the traders $i$ and $j$ share at least one trading day of activity. This means that being $E$ and $E_P$ the sets of edges in the bipartite and in the projected network respectively,
\begin{equation*}
    (i,j) \in E_P \iff \text{for } i, \ j \in A \ \exists t \in B \text{ s.t. } (i,t) \in E \text{ and } (j,t) \in E \ .
\end{equation*}
Since edges in the bipartite system can be of three types according with the states $b$, $s$ and $bs$, in the projected network of traders we can have $9$ types of links: $bb$, $ss$, $bsbs$, $bs$, $sb$, $bbs$, $bsb$, $sbs$, $bss$. For instance, if between $i$ and $j$ there exists a link of the type $bbs$, this means that there is at least one trading day in which $i$ is in state $b$ and $j$ is in state $bs$. If we consider the network in Figure \ref{example_bipartite}, this situation corresponds to traders $i=1$ and $j=2$ since in day $t=1$, $i$ is in state $b$ and $j$ is in state $bs$.

The projected network of traders is a weighted network where weights are given by the number of days for which a given co-occurrence of states between two traders occurs. Let us consider the example above i.e. in the projected network there exists an edge $ (i,j)$ of the type $bbs$. The weight of this link is given by the number of days for which trader $i$ is in state $b$ and $j$ is in state $bs$ i.e. 
\begin{equation*}
    w_{ij} = |H| \text{ where } H \subset B \text{ s.t. } s(i,t) = b \text{ and } s(j,t) = bs \ \forall t \in H.
\end{equation*}
If two traders $(i,j)$ do not share trading days, $w_{ij} = 0$.

Thus, we obtain a multilink weighted graph with $9$ types of link which can be formalized as $G_P = (V_P, E_P, L)$ where $V_P = A$, $L = \{bb, ss, bsbs, bs, sb, bbs, bsb, sbs, bss \}$,
\begin{equation*}
    E_P = \{(w_{ij}, l_{ij}) \ i,j \in V_P, \ l_{ij} \in L \}
\end{equation*} 
and $w_{ij}$ as defined above.

This graph is almost complete and so, before performing the clustering, we identify links which are statistically validated against a null hypothesis. These links should highlight the structure and the organization of the system since their presence cannot be explained by a random assignation of co-occurrences. In doing so, the Statistically Validated Networks (SVN) method \cite{svn1, svn2} is employed.

Let us consider the link $(i,j) \in E_P$. First, we define as $N_i^Q$ the number of days in which trader $i$ is in state $Q \ (b, \ s$ or $bs)$. Analogously, $N_j^R$ is the number of days in which trader $j$ is in state $R \ (b, \ s$ or $bs)$. Then, $N_{ij}^{QR}$ is the number of days in which $i$ is in state $Q$ and $j$ is in state $R$.

The null hypothesis is the random co-occurrence of state $Q$ for investor $i$ and state $R$ for investor $j$. Thus, the probability of observing $X$ days out of $T$ in which the two traders are in the given states, is described by the hypergeometric distribution $H(X|T, N_i^Q, N_j^R)$ and its p-value is defined as
\begin{equation*}
    p(N_{ij}^{QR}) = \mathbb{P}(X \geq N_{ij}^{QR}) =   1 -  \sum_{X=0}^{N_{ij}^{QR} - 1} H(X|T, N_i^Q, N_j^R) \ . 
\end{equation*}
If $p(N_{ij}^{QR})$ is lower than a threshold $p$, the link is validated. 

Concerning the choice of the statistical threshold, it is fundamental to observe that multiple hypothesis tests are being performed. Thus, suited corrections should be employed. According to the Bonferroni correction, the threshold is the usual single-test significance level divided by the number of performed tests i.e. $p = \alpha/N_{test} = 2\alpha/(9N(N-1))$ and we choose $\alpha = 0.01$. Another approach is the so-called False Discovery Rate (FDR), which is less stringent than Bonferroni. On the one hand, FDR is more prone to validate false positives, on the other hand its statistical power is greater thus resulting in the validation of less false negatives. All the p-values are sorted in increasing order i.e. $p_1 \leq p_2 \leq \ldots \leq p_{N_{test}}$; then, the FDR threshold is $p_{\bar{k}}$ where $\bar{k} = \arg\max_{k=1,\ldots,N_{test}} k$ such that $p_{\bar{k}} \leq \bar{k} \alpha/N_{test}$.

Once all tests are performed, the validated projected network of traders is obtained and clustering of traders can be carried out. We adopt the minimalist approach which was proposed in \cite{svn2}: communities of traders are detected on the network which only considers diagonal links i.e. $bb$, $ss$, $bsbs$. As in the previously cited papers, in the empirical analysis described below we find that the number of diagonal links is predominant compared to that of non-diagonal links. Moreover, our goal is to find clusters of investors with similar trading activity around PSEs that are best captured by diagonal links.

\subsubsection{Identifying clusters with Infomap}
Analogously to \cite{svn1}, in order to obtain clusters, the Infomap method for community detection in networks is employed on the validated networks. Infomap is a community detection method. It is an algorithm which minimizes the map equation over possible network partitions \cite{infomap}. It is part of the flow models for community detection indeed, the map equation for a given network partition represents the information cost of a random walker which moves on the partition. Infomap amounts at finding the network partition for which the information cost is minimum. The map equation allows to obtain a result which is less likely to be affected by resolution limit in community detection and for this reason it is considered one of the best methods. From a practical point of view, we relied on its implementation in the Python package \textit{infomap}.


\subsubsection{Characterising clusters}\label{subsubsection_clusters}
Given traders' clusters, their composition can be characterized by relying on the method introduced in \cite{svn1}, analogously to what is done in \cite{svn2,baltakiene}. In particular, we  investigate the over-expression and under-expression of the investors' categories and co-occurrences of states.

The method is again based on the hypergeometric distribution as a benchmark for randomness. We have a total of $N_V$ elements divided in $N_C$ clusters. In order to test whether the attribute $Q$ is over-expressed in the cluster $C$, we compute the p-value
\begin{equation*}
    p_o(N_{CQ}) = 1 -  \sum_{X=0}^{N_{CQ} - 1} H(X|N_V, N_C, N_Q) 
\end{equation*}
where $N_Q$ is the total number of elements in the system with attribute $Q$. If $p_o(N_{CQ})$ is lower than a statistical threshold corrected with Bonferroni or FDR, then the null hypothesis is rejected and we can conclude the attribute $Q$ is over-expressed in cluster $C$.

In a similar way, we test whether an attribute $Q$ is under-expressed in cluster $C$ by comparing the p-value
\begin{equation*}
    p_u(N_{CQ}) = \sum_{X=0}^{N_{CQ}} H(X|N_V, N_C, N_Q) 
\end{equation*}
with a given statistical threshold.

\subsubsection{Identification and classification of potential insider rings.} 

After the identification of the groups of investors, the method is used to identify potential insider rings, i.e. small groups of investors which trade in a synchronized way and in a rewarding position before the PSE. If, for example, the rewarding position is to buy, looking for coordinated suspicious clusters consists in finding groups of investors who are in the $b$ state in the proximity of the PSE.

 In order to characterize more quantitatively clusters' suspicious behavior, some aggregated metrics are considered. The focus is on a time window $\bar{\Delta}$ before the PSE; this could be considered as an average reference period observed by the competent Authorities while investigating insider trading related to the kind of PSE here considered.
 Then, metrics are computed on $\bar{\Delta}$ as averages over each cluster. They are the mean directionality $R_C$ and the mean expected profit $\pi_C$. $R_C$ is the average over the cluster $C$ of the metric $r(i,t)$ (Equation \ref{eqr}).
The mean expected profit $\pi_C$ is defined as the average expected profit of the traders in cluster $C$ computed with respect to the takeover bid share price $p_{TB}$ announced at the time of the PSE:
\begin{equation}\begin{split}\label{exp_profit}
    \pi_C &= \frac{1}{N_C} \sum_{i \in C} \pi_i \text{ \ where}\\
    \pi_i &= p_{TB}\Bigg(\sum_{t \in \bar{\Delta}}{[V_b(i,t) - V_s(i,t)]}\Bigg) - \sum_{t \in \bar{\Delta}}{[A_b(i,t) - A_s(i,t)]} 
\end{split}
\end{equation}
where $N_C$ is the number of traders in cluster $C$, $V_{b/s}(i,t)$ are the volumes (in shares) bought/sold by trader $i$ in day $t$, $A_{b/s}(i,t)$ are the amounts (in Euro) bought/sold by trader $i$ in day $t$. 

\section{Data}\label{sec:data}
\subsection{Transaction reporting database}\label{transaction_data}
The analysis is based on transaction reports collected by Consob for the Italian stocks, according to the directive \href{https://eur-lex.europa.eu/legal-content/IT/ALL/?uri=CELEX:32004L0039&qid=1435044997184}{2014/65} by European Union, also called MiFID II\footnote{In a nutshell, the MiFIDII/MiFIR regime has introduced new regulations for European financial markets and, among them, the transaction reporting obligation that requires investment firms or intermediaries executing transactions in financial instruments to communicate ``complete and accurate details of such trans- actions to the competent authority as quickly as possible, and no later than the close of the following working day''.}
The relevant dataset was built aggregating the daily transactions of all investors operating in any of the Italian stocks, in the period from January 1, 2019 to September 30, 2021. In details, the dataset was built according to the following rules: i) all the information related to the identity of individual investors have been anonymized; ii) with reference to each stock (identified by its ISIN code), each data point keeps a record of:

\begin{enumerate}
    \item anonymous identifier of the investor;
    \item type of investors (household: H, investment firm: IF, legal entity: L); 
    \item trading venue of the operation (Borsa Italiana - MTA, London Stock Exchange - LSE, off-exchange, etc.) for a total of \ldots;
    \item day of the operation;
    \item buy and sell volumes (in shares);
    \item buy and sell Euro volumes;
    \item number of buy and sell contracts;
    \item price of both the first and the last contracts (if there are more than one contract, otherwise they coincide);
    \item minimum and max prices of contracts (if there are more than one contract, otherwise they coincide);
    \item average price of buy (sell) contracts.
\end{enumerate}
In the period covered by the dataset, 2,253,707 investors were observed, operating in 286 Italian stocks. This dataset was recently used in the investigation of the trading behavior of Italian investors during the Covid pandemic (\cite{deriu}).

\subsection{Price sensitive events database}\label{data_PSE}
In addition to the transaction reporting database, a data set containing several price sensitive events (PSEs) was built; such events, obviously public, had all been analysed by the competent Authority with the aim of market abuse detection, by the way of standard analytics methodologies. PSEs are events or a set of circumstances relating to listed companies which, when made public, had a significant impact on the price of the company's shares.

Our focus is on insider dealing in the Italian Stock Exchange. Investors who know in advance when a PSE will occur, can trade in a rewarding manner before the information spreads, thus closing their position after the PSE and making a profit. For instance, if a trader knows a few days before its public announcement that a takeover bid is going to occur for a given stock, they could exploit such information by buying shares of the stock considered. When the takeover bid occurs, the shares' price goes up aligning with the offer price and thus, the informed trader can sell by making a no-risk profit.

PSEs dataset contains a list of 
takeover bids for a number of stocks. As known, a takeover bid is a public offer made by a physical person or a legal entity who is willing to buy other shareholders' shares at a price higher than the stock market value.
As we saw, takeover bids can be exploited by an informed trader by buying before the event. It is worth mentioning that takeover bids have prolonged effects on the market, thus an insider can make a profit even without closing the position immediately after the announcement.

Our data report for each PSE the stock, the type of the event, its date, and the time window for insider trading analysis. This reference period for investigation varies depending on the type of PSE, which leads to different definitions of the time at which an information starts to be considered price sensitive. In Table \ref{resPSE}, the PSEs database is displayed.

\begin{table}[]
    \centering
    \small
    \begin{tabular}{c|c|c}
         Stock  & PSE date & Reference period \\
         \hline
         IMA  & July 28, 2020 & June 29, 2020 - July 28, 2020 \\
         UBI  & Feb 17, 2020 & Jan 16, 2020 - Feb 17, 2020 \\
         PANARIAGROUP  & Mar 31, 2021 & Mar 1, 2021 - Mar 31, 2021 \\
         CARRARO  & Mar 28, 2021 & Jan 4, 2021 - Mar 28, 2021 \\
         MOLMED  & Mar 17, 2020  & Dec 2, 2019 - Mar 17, 2020 \\
    \end{tabular}
    \caption{Price sensitive events database. For each case, the stock name, the type of the event, its date, and the time window for the analysis are reported.} 
    \label{resPSE}
\end{table}

\section{Results}\label{sec:results}
We tested our methods on a set of specific PSEs, namely takeover bids, which are listed in Table \ref{resPSE} together with the considered reference period. For space reason, in the following we present in detail a case study related to the the takeover bid for the Italian stock ``Industria Macchine Automatiche'' (IMA) announced on July 28, 2020. In  the Tables and in the Appendix we present also results obtained by investigating the other PSEs.

The data set related to IMA covers the period January 2, 2019 - February 15, 2021  i.e. $T = 541$ trading days. The starting number of active investors\footnote{An investor is active if he/she executed at least one transaction in the observed time frame. In the text the words investor and trader are used indifferently.} is $26,911$ and the total number of records is $214,122$. Summary statistics of the data set is reported in the top panel of Table \ref{ima_stat0} by referring to the investors' grouping of our database, as explained in Section \ref{transaction_data}. We observe $95.5 \% $ of investors are represented by households and their corresponding exchanged volume amounts at $6.7 \%$ of the total. On the other hand, investment firms which are the $0.5 \%$ of total investors, trade $59.0 \%$ of the total volume. 

\begin{table}[]
    \centering
    \small
    \begin{tabular}{c|c|c|c||c|c|c}
 IMA   & \multicolumn{3}{c||}{Entire set} & \multicolumn{3}{c}{Restricted set} \\
    \hline
         Investor type & N & C & V & N & C & V \\
         \hline 
         Households & $25,705$ & $248,069$ & $21$ & $4,405$ & $151,575$ & $13$  \\
Inv. firms & $145$ & $1,259,860$ & $184$ & $81$ & $1,255,272$ & $181$ \\ 
Legal entities & $1,061$ & $607,650$ & $107$ & $358$ & $581,933$ & $98$ \\ 
\hline
Total & $26,911$ & $2,115,579$ & $312$ & $4,844$ & $1,988,780$ & $292$ \\ 
\multicolumn{7}{c}{}\\
    & \multicolumn{3}{c||}{Entire set} & \multicolumn{3}{c}{Restricted set} \\
    \hline 
         Asset/Total & N & C & V & N & C & V \\
         \hline 
         UBI & $154,080$ & $7,582,854$ &  $18,793$  & $11,977$ & $7,084,384$ & $17,446$ \\ 
         PANARIAGROUP & $6,015$ & $125,416$ & $142$ & $663$ & $91,096$ & $90$\\ 
         CARRARO & $9,019$  & $179,521$ & $202$ & $608$ & $123,643$ &  $125$\\ 
         MOLMED & $17,877$ & $342,703$ & $4,725$ & $2,472$ & $271,364$ & $2,587$\\ 
    \end{tabular}
    \bigskip
    \caption{Top panel. A summary of IMA's data set, before (entire set, used for the k-mean analysis) and after (restricted set, used for the SVN analysis) setting the threshold on trading activity days, is reported. $N$ is the number of traders, $C$ is the contracts' number, $V$ is the sum of bought and sold volume (in millions of shares). Bottom panel. The same (total) quantities are obtained for the other assets involved in a takeover bid analysis. In this second table, the different types of investors are aggregated.}
    \label{ima_stat0}
\end{table}

\begin{figure}
    \centering
    \includegraphics[width=.5\linewidth]{ 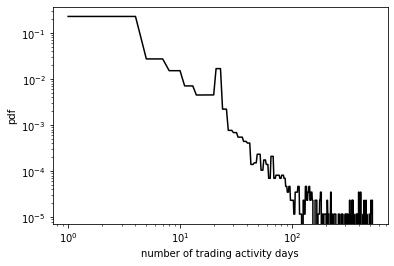}
    \caption{Probability density function (pdf) of the trading activity days for investors in IMA data set. The plot is with log scaling on both axes}
    \label{ima_pdf_tradingdays}
\end{figure}

By looking at the number of days with trading activity for each trader in the investigated period, we observe that this number is small for a wide range of investors. This is shown in Figure \ref{ima_pdf_tradingdays} where the probability density function of investors' trading activity days is displayed. 

In the SVN analysis below, we will consider a threshold on the number of days in which an investor has traded, then obtaining the restricted set of the most active investors in the market. Summary statistics of this restricted data set are reported also in the top panel of Table \ref{ima_stat0}. Finally, in the bottom panel of Table \ref{ima_stat0}, the same statistics, but aggregated over all types of investors, are shown for the other assets involved in a takeover bid analysis.


\subsection{K-means results}\label{sec_kmeans_res}

The {\it k-means} clustering method described in Section \ref{sec_kmeans} is applied by defining: (i) $t_S$ as the date of the PSE, namely July 28, 2020, (ii) the length of the time window $\vert\Delta\vert$ as the duration of the reference period shown in Table \ref{resPSE}, i.e. one business month of $20$ days, and (iii) January 2, 2020 as initial time $t_1$. We consider a rolling window $\Delta$ starting from January 2020, then shifted week by week until the date of the PSE, for a total of $27$ time windows. Within each time window, the features characterizing the trading behavior of an investor are computed as described above.

\begin{figure}
    \centering
    \includegraphics[width=.45\linewidth]{ 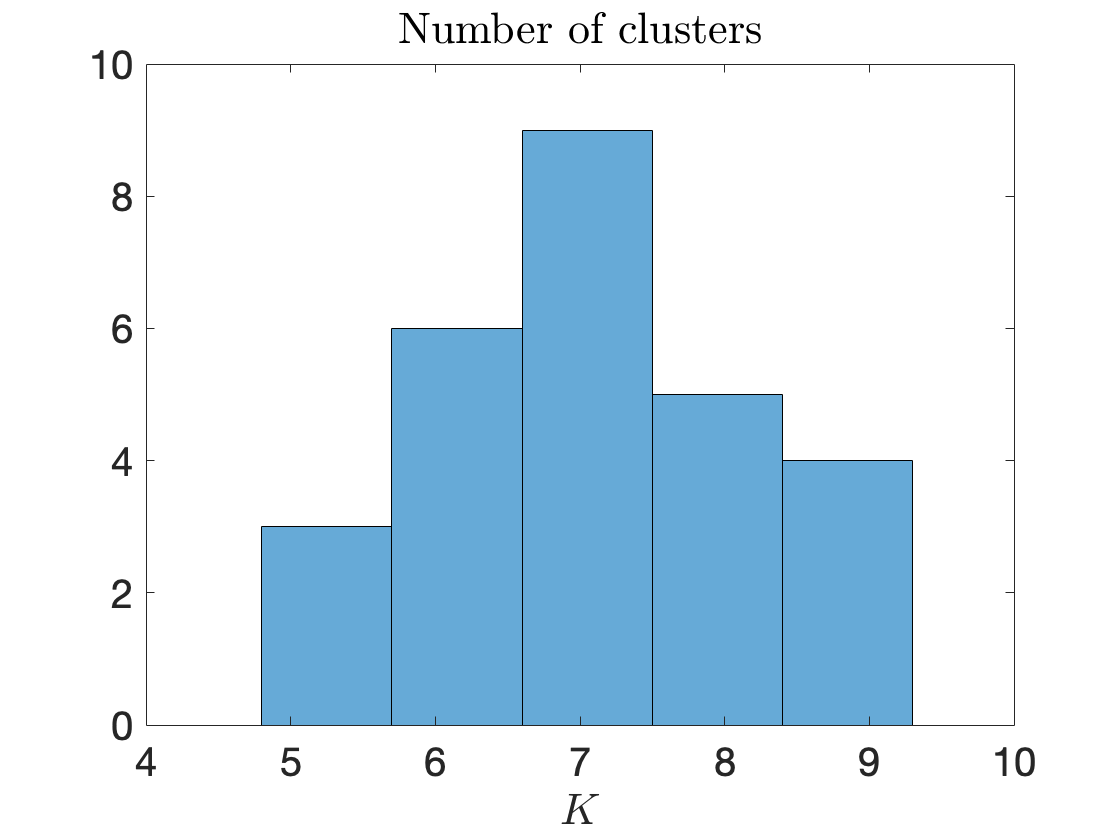}
    \includegraphics[width=.45\linewidth]{ 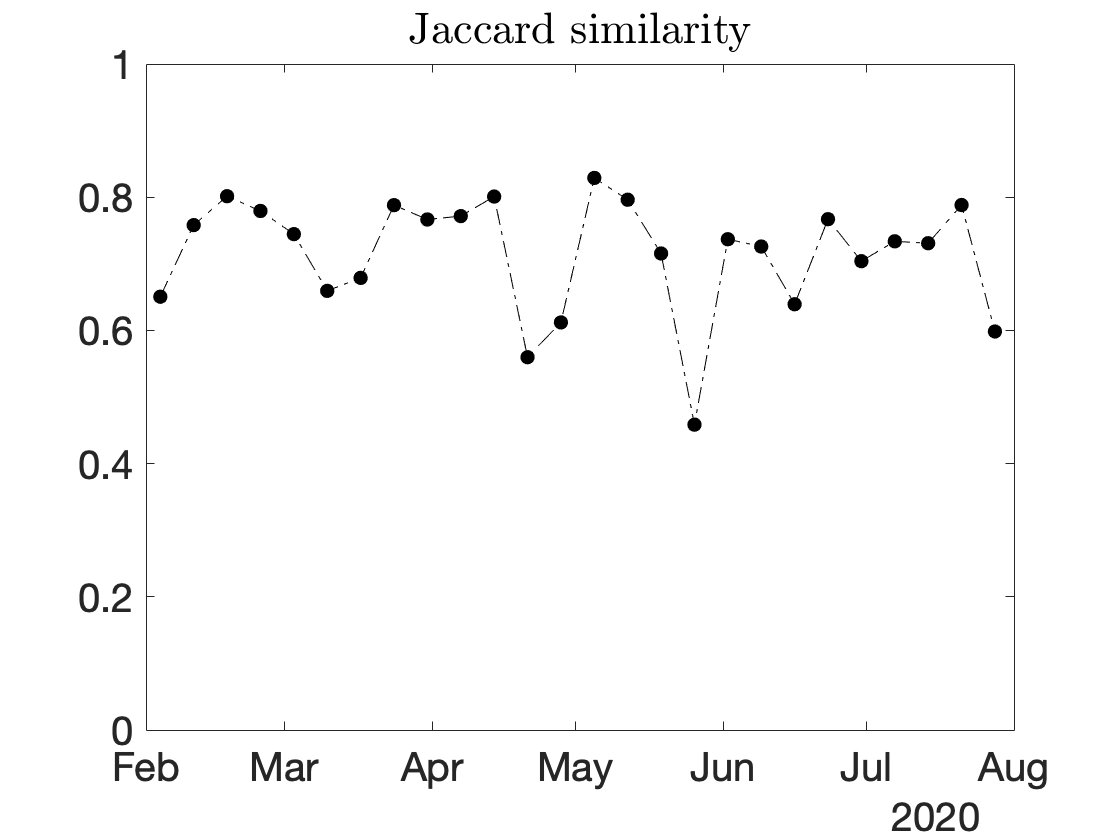}
    \includegraphics[width=.45\linewidth]{ 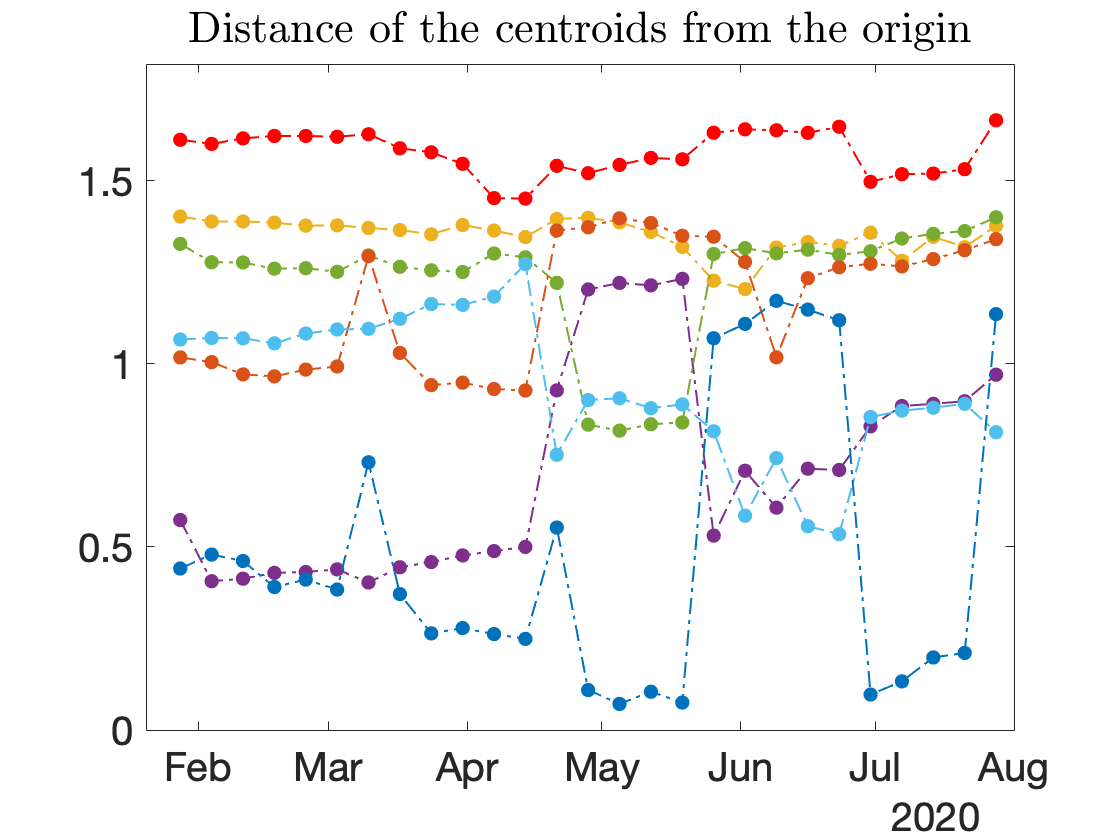}
    \caption{Distribution of optimal number of clusters $K$ (left panel). Jaccard similarity between two clusters in a row for IMA (middle panel). Evolution of Euclidean distances of each cluster's centroid from the origin (right panel).}
    \label{ima_kmeans}
\end{figure}

\begin{figure}
    \centering
    \includegraphics[width=1.\linewidth]{ 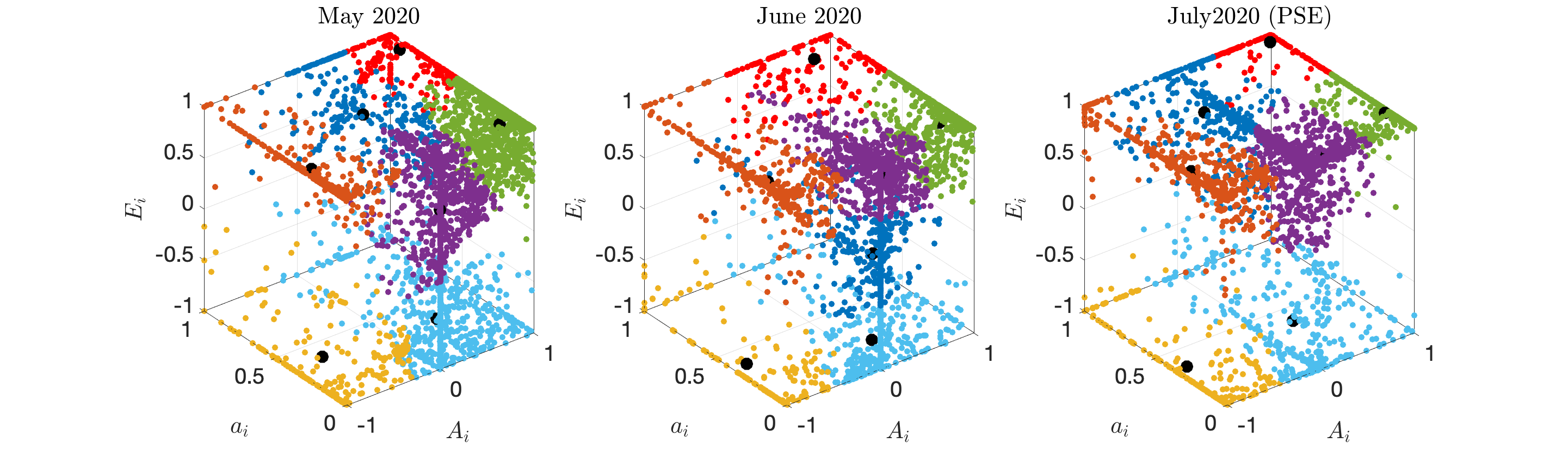}
    \caption{Inferred clusters for IMA ($K=7$), for three different months.}
    \label{ima_kmeans_clusters}
\end{figure}

Before running the algorithm we find the optimal number of clusters $K$ in each time window. The distribution of optimal $K$s is shown in the left panel of Figure \ref{ima_kmeans}. We select a single hyperparameter $K$ for each time window as the rounded mean of the distribution, obtaining $K=7$. We finally consider the {\it dynamic clustering} approach, achieving a stable description of the clusters in which we partition the feature space. This is confirmed by the high value of the Jaccard similarity coefficient, as shown in the middle panel of Figure \ref{ima_kmeans}. In fact, the positions of the centroids in the feature space is almost stable in the whole period, see the right panel of Figure \ref{ima_kmeans}. 

A pictorial representation of the clusters in the Euclidean space at three different months (May, June, July) is shown in Figure \ref{ima_kmeans_clusters}, where each cluster is identified by a different color. We can notice that almost all clusters are stable from one month to the next one, with the exception of the blue and the purple ones. Moving from May to June and then from June to July, the optimal partition shows some switching behavior depending on the distribution of blue and purple features at different months. 

\begin{table}[]
    \centering
    \small
    \begin{tabular}{c|c|c|c|c|c}
         Stock & K & Soft disc. & Hard disc. & $\%$ disc. (red) & avg. $\%$ disc. (others) \\
         \hline
         IMA & $7$  & $66$  & $237$  & $0.86^{*}$ & $0.50$  \\ \hline
         CARRARO &  $6$ & $43$ & $388$  & $0.98^{**}$ & $0.68$ \\
         MOLMED &  $7$ & $25$ & $259$  & $0.89^{**}$ & $0.63$  \\
         PANARIAGROUP & $8$ & $5$ & $27$  & $0.82^{*}$ & $0.60$ \\
         UBI & $7$  & $204$ & $1304$  & $0.85^{**}$ & $0.67$ \\
    \end{tabular}
    \caption{Number of soft and hard discontinuous traders in the (red) suspect cluster for IMA (and other takeover bids). Percentage of discontinuous (soft + hard) traders in the suspect cluster, which is statistically significant according to the Chi-squared test with level of significance $5\%$ (*) or $1\%$ (**) when compared to other clusters, as described in the main text).}
    \label{disc_frac}
\end{table}

Following the discussion in Section \ref{sec_kmeans}, the cluster more likely to be involved in insider activity is the red one, since it is the one whose centroid is closest to $\bm{1}$. However, being in the red cluster is not necessarily an indication of insider activity, since it is necessary to investigate whether an investor belonged to it also in the reference period or displayed a discontinuous behavior.

For IMA there are $66$ soft discontinuous and $237$ hard discontinuous traders in the red cluster, summing to $303$ discontinuous traders over a total of $379$ investors in the red cluster, see Table \ref{disc_frac}. The question is now whether the fraction of discontinuous traders is statistically significant when compared with other clusters or not. In other words, the crucial point is to know if the cluster with rewarding position is somehow anomalous when looking at the discontinuity of trading behavior of investors. A pairwise comparison can be performed in a statistical fashion by using the $\chi^2$-test. Given two clusters, whose investors are labeled as continuous or discontinuous traders, we can test the null hypothesis that the labels have equal probability in the two clusters (e.g. if the fraction of discontinuous traders in the red cluster is consistent with the fraction of discontinuous traders in another cluster) by considering the following statistic
\begin{equation*}\label{chi2}
    \chi^2 = \sum_{\ell\in\{C,D\}}\frac{(n^{(1)}_\ell-n^{(2)}_\ell)^2}{n^{(1)}_\ell+n^{(2)}_\ell}
\end{equation*}
where $\ell$ indicates the label, i.e. continuous $C$ or discontinuous $D$, and $n^{(1)}_\ell$ ($n^{(2)}_\ell$) is the number of investors with label $\ell$ in the first (second) cluster. Under the null hypothesis, the test statistic (\ref{chi2}) is $\chi^2$-distributed with one degree of freedom. When the p-value associated with the test statistic is below a given confidence level, e.g. $5\%$ or $1\%$, the null hypothesis is rejected and it is possible to conclude that two clusters are statistically different. In the case under investigation, the comparison between the red cluster and any other one is done by performing $K-1$ $\chi^2$-test, i.e. red vs. any other color. The results lead to the conclusion that the red cluster is statistically different from the others if the null hypothesis is always rejected. For IMA and PANARIAGROUP this is true at $5\%$ confidence level while for the others at $1\%$ level. In particular, the fraction of discontinuous traders in the red cluster is always larger than others. This result is summarized in Table \ref{disc_frac}, where the percentage of discontinuous traders in the red cluster and the average percentage in other clusters is displayed. 

\begin{table}[]
\centering
\small
\begin{tabular}{l | l | c | c | c | c | c }
Ranking & Anonymous ID & Type & Score & Shares & Directionality & Exp. Profit (\texteuro)  \\
\hline
1& 257853 & L & 1  & 52000  & 1 & 443,947  \\ 
2&  783031 & L  & 1  & 23535  & 0.49 & 320,457 \\ 
3 &  1664331 & L & 1  & 19500  & 1 & 277,625\\
4 &  1139263 & L  & 1 & 17250 & 0.81 & 241,132 \\
5 &  9280051 & L & 1 & 16700 & 1 & 138,132\\
6 &  9276483 & L & 1  & 10500  & 1 & 94,727 \\
7 & 9249321 & L & 1 & 9000 & 1 &  101,782 \\
8 &   2193864 & H & 1 & 5700 & 1 & 94,006  \\
9 & 1564905 & H  & 1 & 4153 & 1 & 62,729  \\
10 &  9249741 & L & 1 & 4001 & 1 & 51,084   \\
11 & 9253185  & L & 1 & 4000 & 1 & 44,778 \\
12 &  208123 & L  & 1 &  3559 & 0.31 & 44,961 \\
13 & 9253505  & H & 1 & 3500  & 1 & 32,200  \\
14 & 9253442 & H & 1 &  3245 & 1 & 35,786 \\
15 & 9239385 & H & 1 &  3000 & 1 & 45,498 \\
\vdots &  \vdots & \vdots  &\vdots  & \vdots & \vdots & \vdots \\
225 & 948008 & H  & 0.99 & 100 & 1 & 1,671\\
226 & 7882312 & H  & 0.98 & 940 & 1 & 14,202 \\
\vdots &  \vdots & \vdots  &\vdots  & \vdots & \vdots & \vdots \\
303 & 135723 &  L & 0.55 & 3000 & 0.22 & 53,367
\end{tabular}
\caption{Ranking of the discontinuous traders in the red cluster for IMA according to the score function defined in the main text. The number of shares bought in the reference period, together with the directionality of trading and the expected profit of the trading in the reference period (by closing the position marked to the takeover bid share price) are shown.}
\label{tab_ranking}
\end{table}

Once the set of discontinuous investors has been obtained, we can sort them according to the score $s_i$ defined in Section \ref{sec_kmeans}, which is inversely related to the distance from the $\bm{1}$ point in the cube.
The ranking for IMA is shown in Table \ref{tab_ranking}. Notice that there are $224$ discontinuous traders in the highest rewarding position (score equal to one) over a total of $303$ investors. This subset is then ranked according to the number of shares bought within the reference period. The directionality of the operations, as defined in Eq. (\ref{eqr}), and the expected profit, as defined in Eq. (\ref{exp_profit}) and with respect to the takeover bid share price $p_{TB} = 68.0$ \texteuro, \ are also shown. 

This kind of ranking is the final output of the methodology introduced here and needs to be interpreted as support to the investigation for insider trading by the regulator.

\subsection{SVN results}\label{sec:svnresults}

We now describe the application of the SVN method to the IMA takeover bid to identify small groups of potential insiders. Since the method is based on a statistical set, analogously to what is done in \cite{svn1}, we restrict the analysis to the investors who traded at least $8$ days in the investigated time period. Setting this threshold allows us to reduce the statistical uncertainty which is typical of rare events.
The number of traders under investigation drops to $N = 4,844$ as shown in Table \ref{ima_stat0} together with a summary of this restricted data set. From this table, it is evident that the investors in the restricted data set i.e. the ones who are active in at least $8$ days, trade $93.6 \%$ of the total exchanged volume.

Given this data restriction, we proceed as explained in subsection \ref{subsection_method}: states and the projected network of traders are obtained, then links are statistically validated with both Bonferroni and FDR corrections and finally, clustering is performed via Infomap.

The Bonferroni threshold turns out to be equal to about $9.47 \cdot 10^{-11}$ while the FDR threshold is about $3.39 \cdot 10^{-4}$.  Table \ref{ima_edges} shows the number of validated edges we obtained with the two corrections. With both corrections, the majority of validated links belong to the $bb$ and $ss$ types and are about $3$ millions. This represents a consistent reduction in the number of possible edges in the projected network of traders which amounts at $N(N-1)/2*9 \simeq 10^8$. However, as we will see in the following, this number is larger than the one obtained for assets with similar number of investigated records and this could be due to the fact than in the IMA case there are stronger signals of synchronization.

\begin{table}[]
    \centering
    \small
    \begin{tabular}{c|c|c}
         Edge type & Bonferroni & FDR \\
         \hline
         $bb$ & $1,320,213 \ (42.21)$ & $1,468,510$ $( 41.00)$ \\
         $ss$ & $1,807,170 \ (57.78)$ & $2,042,470$ $( 57.02)$ \\\
         $bsbs$ & $41 \ (< 0.01)$ & $287$ $(< 0.01)$\\
         $bs$ & $30 \ (< 0.01)$ & $33,204$ $(0.93)$\\
         $sb$ & $49 \ (< 0.01)$ & $28,541$ $(0.80)$ \\
         $bbs$ & $31  \ (< 0.01)$ & $4,314$ $(0.12)$\\
         $bsb$ & $43 \ (< 0.01)$ & $2,562$ $(0.07)$ \\
         $sbs$ & $26 \ (< 0.01)$ & $1,159$ $(0.03)$\\
         $bss$ & $53 \ (< 0.01)$ & $939$ $(0.02)$\\
         \hline
         Total & $3,127,656
$ & $3,581,986$
    \end{tabular}
    \caption{Numbers of the different types of links in the IMA's SVN obtained with the Bonferroni and FDR corrections. Numbers in parenthesis are the corresponding percentage values.}
    \label{ima_edges}
\end{table}

Focusing on the diagonal links i.e. $bb$, $ss$ and $bsbs$, and on the resulting network, the clustering is performed. With the Bonferroni correction, the SVN is made up of $2,434$ non-isolated nodes and $3,127,424$ edges. 
With FDR instead, the SVN is made up of $4,673$ non-isolated nodes and $3,511,267$ edges. 
See Appendix \ref{appendix_details_ima} for further details about SVN traders' type. When Infomap is run for both the Bonferroni and FDR SVN we obtain $69$ and $13$ clusters, respectively. In Appendix \ref{appendix_details_ima}, tables with summary statistics about the most populated clusters are reported together with an analysis of the relation between the FDR and the Bonferroni network.



\begin{figure}
    \centering
    \includegraphics[width = 12cm]{ 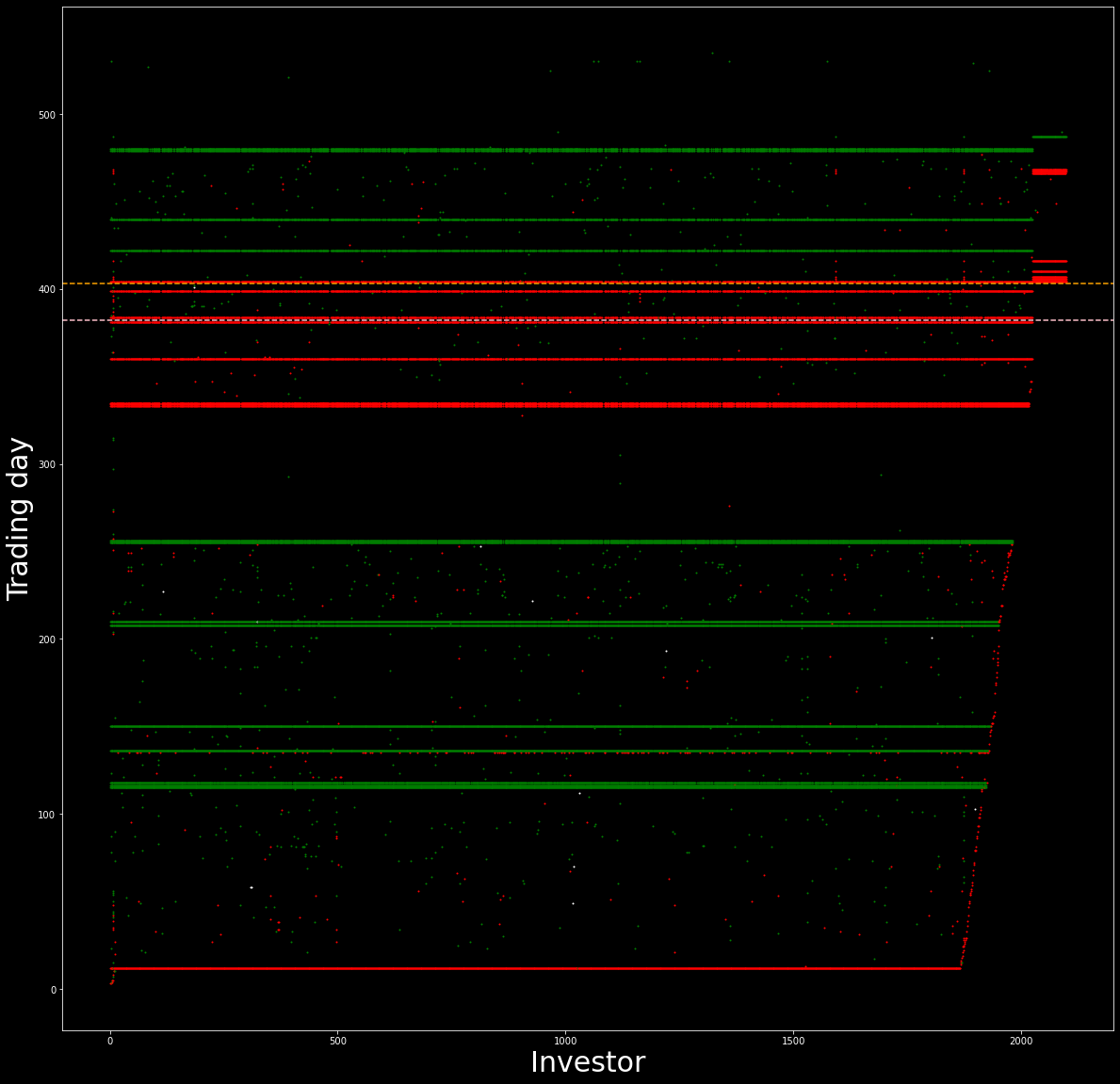}
    \caption{Graphical representation of traders' activity in the cluster $1$ obtained by running Infomap on the Bonferroni SVN $bb$-$ss$-$bsbs$. The $x$-axis represents investors, the $y$-axis trading days. Black points correspond to no-activity, red to $b$ state, green to $s$ state and white to $bs$ state. The dotted orange horizontal line marks the PSE and the dotted pink horizontal line marks the beginning of the time window $\bar{\Delta}$.}
    \label{states1_ima_Bonferroni}
\end{figure}

\begin{figure}
    \centering
    \includegraphics[width = 12cm]{ 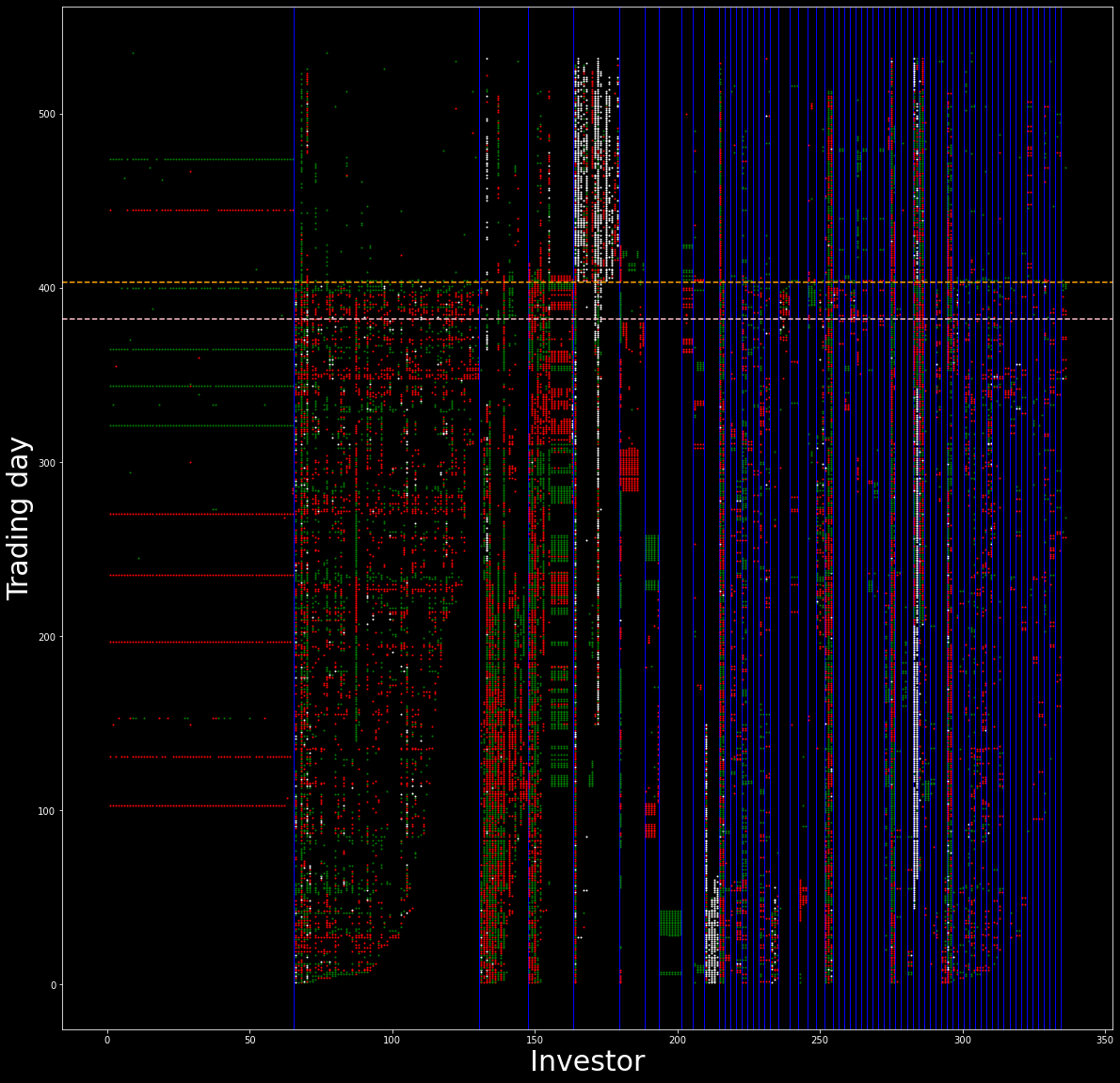}
        \caption{Graphical representation of traders' activity in the cluster $2-63$ obtained by running Infomap on the Bonferroni SVN $bb$-$ss$-$bsbs$. The $x$-axis represents investors, the $y$-axis trading days. Black points correspond to no-activity, red to $b$ state, green to $s$ state and white to $bs$ state. Vertical light blue lines separate the clusters, the dotted orange horizontal line marks the PSE, the dotted pink horizontal line marks the beginning of the time window $\bar{\Delta}$.}
    \label{states2-end_ima_Bonferroni}
\end{figure}

Once the clusters have been obtained, we display the activity of the investors which do not correspond to isolated nodes as suggested in \cite{svn1}, providing an extremely useful representation of coordinated behaviors of investors around the PSE. Figures \ref{states1_ima_Bonferroni} and \ref{states2-end_ima_Bonferroni} show the trading activity of non-isolated nodes in the Bonferroni\footnote{In Appendix \ref{appendix_details_ima}, the corresponding activity plots for the FDR SVN are reported.} SVN. The $x$-axis represents investors and the $y$-axis trading days. Black points correspond to no-activity, red to $b$ state, green to $s$ state and white to $bs$ state. Therefore, each vertical sequence of points represents the activity over time of a given trader. Vertical light blue lines separate the clusters obtained via Infomap and the dotted orange horizontal line marks the date of the PSE i.e. the takeover bid of July 28, 2020. 

Figure \ref{states1_ima_Bonferroni} shows that the largest identified cluster is composed by more than $2000$ investors, most of them being households, and displays an extremely synchronized behavior. This coordinated behavior is likely explained by investors with portfolios managed by the same entity. In Figure \ref{states2-end_ima_Bonferroni} we observe all the other clusters and it is clear that the degree of trading synchronization inside each of them is very high, displaying, also for this database, the amazing capability of SVN to identify small synchronized clusters.

\begin{table}[]
    \centering
    \small
    \begin{tabular}{c|c|c|c|c|c}
         Cluster & Traders & Traders active in $\bar{\Delta}$ & Traders type & $\pi_c$ (\texteuro)& $R_C$  \\
         \hline
         $33$ & $2$& $1$ & L &$893,456$ &$1.00$\\
         $10$ &$4$ &$4$ & L &  $547,921$ & $1.00$\\
         $31$ & $2$& $1$ &L & $155,805$ & $1.00$ \\
         $62$& $2$& $2$&L & $32,202$ & $1.00$\\
         $63$& $2$& $2$ &IF, L & $2,526$ & $1.00$\\
         $55$ &$2$ &$1$& H & $860$ & $1.00$ \\
         $67$ &$2$ &$2$ & H & $725$& $1.00$\\
         $51$ &$2$ &$2$ &H & $692$ & $1.00$\\
         $59$& $2$& $2$ & H & $78$ &$1.00$\\
         $30$ & $2$ & $2$ & H & $66$ &$1.00$\\
         $37$ &$2$& $2$ & H & $53$& $1.00$ \\
         $34$ & $2$ & $2$ & H & $40$ & $1.00$\\
         $17$ &$2$ &$2$ & L &  $1,584,498$ & $0.99$\\
         $1$ & $2098$ & $1635$& H (99.33 \%) & $145$ & $0.92$\\
         $29$ & $2$ & $2$ & L & $259,379$ & $0.91$\\
         $64$& $2$ &$2$ & L & $92,214$ & $0.88$\\
         $18$ & $2$& $2$ & H & $1,287$ & $0.80$
    \end{tabular}
    \caption{Results related to clusters of the Bonferroni SVN with mean directionality $R_C > 0.5$ are reported. For each cluster, the number of its total traders and the ones active in the time window $\bar{\Delta}$ are specified. For the traders active in $\bar{\Delta}$, their type (H = household, L = legal entity, IF = investment firm) is summarised and then, mean quantities about the cluster - that are the mean expected profit $\pi_C$ in euro and the mean directionality $R_C$ - are shown.}
    \label{clusters_analysis_ima_Bonferroni}
\end{table}

\begin{figure}
    \centering
    \includegraphics[width = 10cm]{ 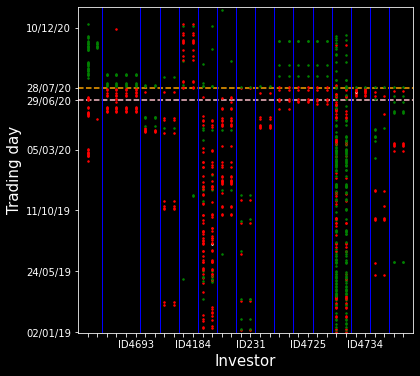}
        \caption{Graphical representation of traders' activity in clusters with mean directionality $R_C  > 0.5$ (except cluster $1$) obtained by running Infomap on the Bonferroni SVN $bb$-$ss$-$bsbs$. The $x$-axis represents investors, the $y$-axis trading days. Black points correspond to no-activity, red to $b$ state, green to $s$ state and white to $bs$ state. Vertical light blue lines separate the clusters, the dotted orange horizontal line marks the PSE and the dotted pink horizontal line marks the beginning of the time window $\bar{\Delta}$.}
    \label{states_clusters_avgdirgreaterthan0_5except1_ima_Bonferroni}
\end{figure}

\begin{table}[]
    \centering
    \small
    \begin{tabular}{c|c|c}
          & \multicolumn{2}{c}{Number of traders in clusters with $R_C \geq  0.9$}  \\
         Stock & Bonferroni & FDR \\
         \hline
         IMA &  $1,662$ & $0$\\
         UBI & $102$ & $28$\\
         PANARIAGROUP & $1$ & $2$\\
         CARRARO &  $0$ & $0$\\
         MOLMED & $3$ & $1$\\
    \end{tabular}
    \caption{Summary table about the number of traders in clusters with mean directionality $R_C \geq 0.9$. The results related to the takeover bids under investigation are reported for both the Bonferroni and the FDR correction.}
    \label{allOPA_SVN_suspected}
\end{table}

We now turn to the use of SVN for the identification of potential insider rings. For the IMA case,  the reference period $\bar{\Delta}$ corresponds to the period from June 29, 2020 to the PSE and the takeover bid share price, used to compute the mean expected profit $\pi_C$, is $p_{TB} = 68.0$ \texteuro.
To identify suspect clusters, we compute he mean directionality $R_C$ and $\pi_C$ for the clusters of both the Bonferroni and the FDR SVN. Table \ref{clusters_analysis_ima_Bonferroni}, reports the results for the Bonferroni clusters with mean directionality greater than $0.5$. Most of these clusters are made up of a couple of traders and in three cases, only one of them is active in the reference period $\bar{\Delta}$. Notice that also the large cluster of Fig. \ref{states1_ima_Bonferroni} is present in the table. Figure \ref{states_clusters_avgdirgreaterthan0_5except1_ima_Bonferroni} shows the activity of the small clusters of Table \ref{clusters_analysis_ima_Bonferroni}. The buying coordinated behavior of these traders in the time window $\bar{\Delta}$ is evident. Several clusters in the figure were essentially inactive in the months preceding the PSE and started trading in very coordinated way with a rewarding position (buy in the IMA case) in the vicinity of the PSE and then closed the position after it. This suggests a suspicious behavior that requires further investigations by the competent Authority.

When we use FDR instead of Bonferroni correction, we find that the obtained clusters have typically a lower directionality and this might be related to the larger number of false positives detected by FDR. As Figure \ref{IMA_bonferroni_vs_fdr} of the Appendix shoes, the FDR cluster number $1$ contains the majority of non-isolated nodes in the Bonferroni SVN.
This issue concerning the FDR correction is also present for other the other investigated PSEs. Table \ref{allOPA_SVN_suspected} shows the number of traders in clusters with mean directionality greater than or equal to $0.9$ for the $5$ takeover bids. It is observed that employing the FDR correction in the IMA case leads to a catastrophic reduction in the number of traders in clusters with high directionality. This reduction is about the $27$ \% for UBI. Also for MOLMED the FDR captures less traders in clusters with high directionality, even if the difference is just of a couple of units. On the other hand, results are unchanged for CARRARO while, for PANARIAGROUP, the FDR correction is more effective at detecting traders in suspicious clusters but, similarly to MOLMED, the difference is just of one unit. 

\begin{figure}
    \centering
    \begin{subfigure}{.5\textwidth}
    \centering
    \includegraphics[width=.75\linewidth]{ 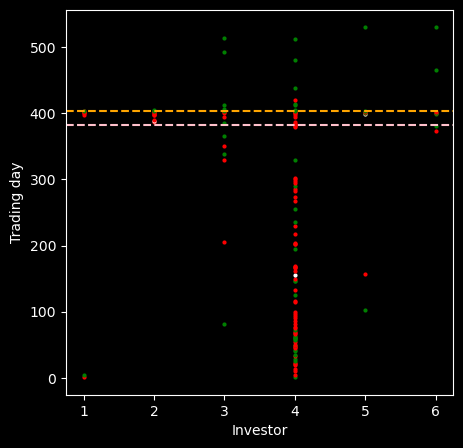}
    \caption{PG\_379532}
    \end{subfigure}%
    \begin{subfigure}{.5\textwidth}
    \centering
    \includegraphics[width=.75\linewidth]{ 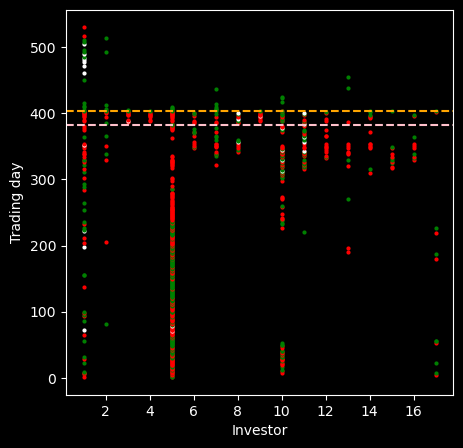}
    \caption{PG\_18731}
    \end{subfigure}
    \begin{subfigure}{.5\textwidth}
    \centering
    \includegraphics[width=.75\linewidth]{ 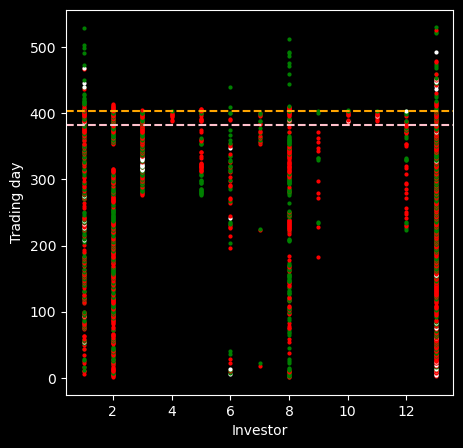}
    \caption{PG\_7877}
    \end{subfigure}%
    \begin{subfigure}{.5\textwidth}
    \centering
    \includegraphics[width=.75\linewidth]{ 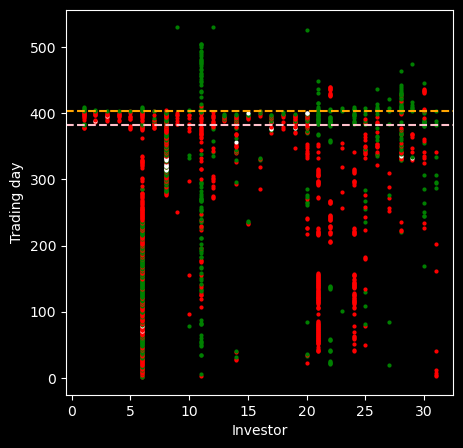}
    \caption{PG\_544592}
    \end{subfigure}
    \begin{subfigure}{.5\textwidth}
    \centering
    \includegraphics[width=.75\linewidth]{ 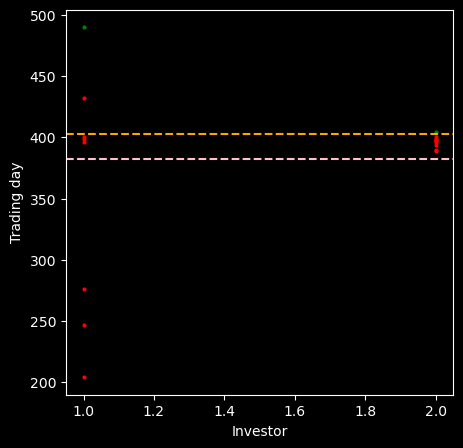}
    \caption{PG\_77838}
    \end{subfigure}
    \caption{Each sub figure is a graphical representation of the trading activity of a suspected node (the first from the left) and its first neighbors in the FDR SVN $bb$-$ss$-$bsbs$. The $x$-axis represents investors, the $y$-axis trading days. Black points correspond to no-activity, red to $b$ state, green to $s$ state and white to $bs$ state. The dotted orange horizontal line marks the PSE and the dotted pink horizontal line marks the beginning of the time window $\bar{\Delta}$. The anonymous IDs of the suspected traders is reported below each sub figure.}
    \label{ima_first_neig}
\end{figure}

The richer nature of the FDR SVN can however be determinant when we would like to use the SVN method in a kind of \textit{human-in-the-loop} manner. Let us suppose that another method for market abuse detection determines that an investor has a suspicious behavior. Then, it is possible to identify investors who are coordinated with the suspicious one in their trading actions, by focusing on the first neighbors in the validated projected network of investors. 
As an example, we will consider the suspicious investors identified according to the {\it k-means}-based methodology in the IMA case. These are isolated nodes in the Bonferroni SVN, while they are not in the FDR SVN.
In Figure \ref{ima_first_neig}, each sub figure shows the trading activity of a suspected trader according to k-means (the one most on the left) 
and of its first neighbors in the FDR SVN. It is clear that several first neighbors exhibit suspicious trading behavior around the takeover bid, but they were not identified by the k-means method.  A more precise analysis is out of the scope of the paper and it is left for future research. 

\begin{figure}
    \centering
    \begin{subfigure}{.5\textwidth}
    \centering
    \includegraphics[width=.8\linewidth]{ 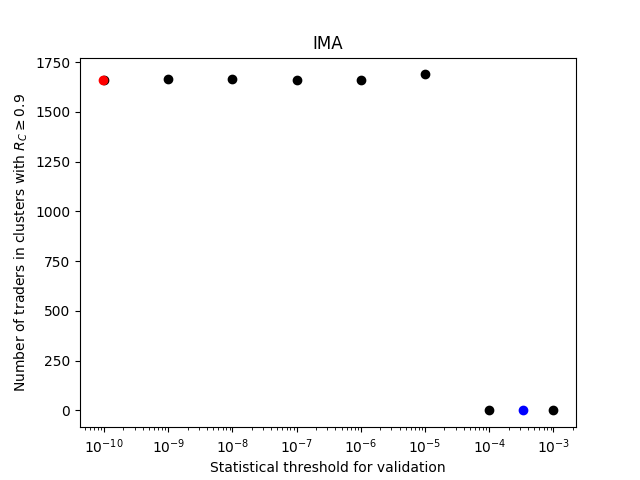}
    \end{subfigure}%
    \begin{subfigure}{.5\textwidth}
    \centering
    \includegraphics[width=.8\linewidth]{ 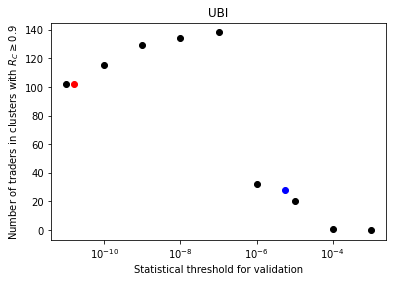}
    \end{subfigure}
    \begin{subfigure}{.5\textwidth}
    \centering
    \includegraphics[width=.8\linewidth]{ 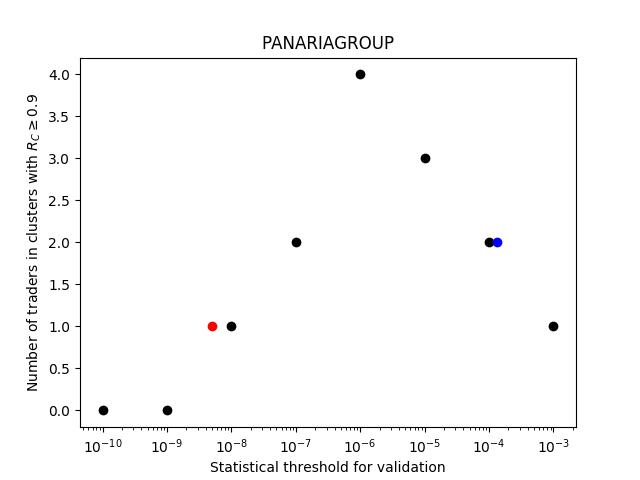}
    \end{subfigure}%
    \begin{subfigure}{.5\textwidth}
    \centering
    \includegraphics[width=.8\linewidth]{ 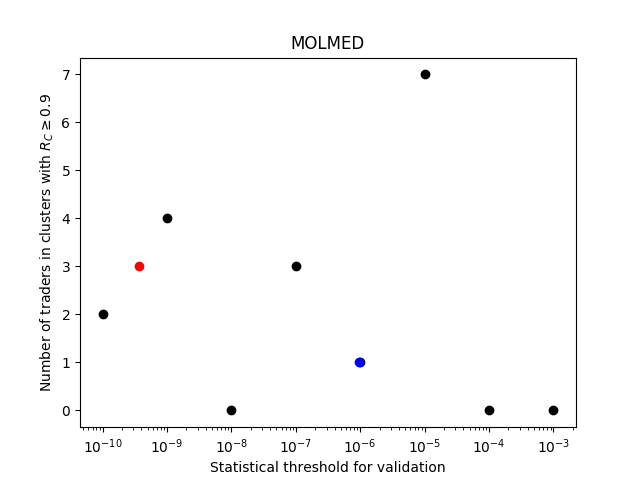}
    \end{subfigure}
    \caption{Each sub figure represents the number of traders in clusters with mean directionality $R_C \geq  0.9$ as a function of the statistical threshold employed for edges' validation. The red and the blue point corresponds to the Bonferroni and FDR correction respectively.}
    \label{threshold_for_validation_analysis}
\end{figure}

As seen above, the choice of the correction used for multiple hypothesis tests is fundamental and can lead to different conclusions about candidate traders as insiders. Then, an option would be treating the statistical threshold for edge validation in our projected network of traders, as a parameter to be optimized. The optimal value of this threshold corresponds to the maximum number of traders in clusters with mean directionality greater than or equal to $0.9$. In Figure \ref{threshold_for_validation_analysis}, this analysis is carried out for IMA, UBI, PANARIAGROUP and MOLMED. The trend observed for IMA and UBI is similar and the maximum is achieved for a statistical threshold equal to $10^{-5}$ and  $10^{-7}$ respectively. This choice would amount at obtaining for IMA, $27$ more traders in suspected clusters than the Bonferroni correction and $36$ for UBI. For PANARIAGROUP, $10^{-6}$ leads to the maximum value $4$. MOLMED instead, has an optimal threshold equal to $10^{-5}$, which leads to $7$ traders in suspected clusters. It is worth noting that MOLMED and PANARIAGROUP are asset with less records than IMA and UBI, as shown by Table \ref{ima_stat0}, and so, this could also explain their fluctuating behavior observed in Figure \ref{threshold_for_validation_analysis}. These results confirm how the choice of an optimal statistical threshold for validation should be carried out case-by-case.

Notice that, in order to check the robustness of the SVN method for market abuse detection, the clustering is performed also with an entropy-based approach introduced in \cite{saracco2017}. The Jaccard similarity \cite{tibi} between couples of clusters obtained with the two procedures is computed and good agreement is found, as it is shown in the results reported in Appendix \ref{appendix_bicm}.


\subsection{Discussion on methods} The \textit{k-means} and the SVN based approaches for market abuse detection are two methods with different purposes: the former aims at capturing discontinuities in the trading operations of single investors (although taking into account the trading behavior of his/her peers). Instead the latter focuses on the identification of coordinated suspicious behavior of groups of investors. Consequently, they detect different anomalies.

Let us focus on IMA again and compare the traders identified as suspicious by the two methods. The \textit{k-means} identified a total of $303$ discontinuous investors, divided in $237$ \textit{hard} and $66$ \textit{soft}. On the other hand, the SVN Bonferroni approach detected $1,662$ traders in clusters with mean directionality greater than or equal to $0.9$. The overlap is very small, resulting in $4$ traders, $2$ households and $2$ legal entities. The two households are \textit{soft} discontinuous traders and they are part of the cluster number $1$ in the SVN Bonferroni. They are both characterized by a strong directionality ($1$ and $0.99$) but different levels of expected profit ($1380$ \texteuro \ and $95,619$ \texteuro). The two legal entities are instead \textit{hard} discontinuous traders with very high expected profits ($277,625$  \texteuro \ and $241,132$ \texteuro) and directionality ($1$ and $0.81$); they are traders number $3$ and $4$ in Table \ref{tab_ranking}. Interestingly, they form a micro cluster of $2$ highly synchronized traders, that is the cluster number $29$ in the SVN Bonferroni.

This comparison highlights how the two methods provide different but complementary results. Given the complexity of the problem, it has to be tackled with an approach that captures several aspects at the same time. Therefore, the lacking of overlapping between the results of the two methods needs to be considered as strength of the proposed approach.


\section{Conclusions}\label{sec:conclusions}

This paper proposes the use of two unsupervised clustering methods for the identification of potential suspects of insider trading in the vicinity of a PSE, such as a takeover bid. The first method clusters the investors in a space of three features and identifies as potential suspects those investors who display a discontinuous trading behavior with respect to their past activity and in a rewarding direction with respect to the PSE. The second method aims at detecting small groups of investors which trade in a synchronized and rewarding way in the vicinity of a PSE, pointing at possible insider rings and collusion in the insider trading activity. The two methods are complementary and indeed the overlap between the potential suspects found by them is small. In our opinion this is an advantage, since they focus on two different aspects of insider trading activity. Moreover, as shown at the end of Section \ref{sec:svnresults}, they can be used jointly to identify investors which are not identified by each method individually. This approach based on the identification of neighbors of suspect investors in the FDR network can also be used by considering suspects obtained with other methods (for example with the traditional supervising approach) rather than with the suspects from k-means.

There are several extensions of this work which we can foresee. First, it would be interesting to extend our analysis to other PSE, such as, for example, Accelerated Bookbuilds or corporate news releases. Second, in the k-means approach we have considered three specific features, which have been chosen both because they are financially relevant for the problem under investigation (insider trading) and because they allow to represent the investors in a three dimensional space. However, it is clear that other features could be added to the clustering analyses obtaining a richer characterization of the investors' population and thus a more precise identification of discontinuous behavior. Third, concerning the SVN approach, we used a trading day to define synchronous trading and for this reason we defined the trading state on a daily level. However, synchronous behavior can occur on longer (or shorter) time scales and one can easily extend our methodology in this direction. Finally the SVN method could shed light on the reference time to monitor suspicious trading activity around a PSE. 

Identification of insider trading is a complicated activity which requires the analysis of large and complex datasets, especially if one wants to consider the activity of potential insiders, not on an individual basis, but through a comparison with the behavior of the whole population of investors. Our proposed methodologies provide two contributions in this direction and we believe they might be very useful in the monitoring activity of supervising authorities. As clearly reported, such methodologies do not constitute any official process of the Consob but represent an in-depth analysis tool to be used when certain investigative conditions are met.


\section*{Acknowledgments}

We wish to thank Sandro Leocata, Francesco Gigante, Carlo Martinoli, Stefania La Civita, Alessio Sanfilippo, Vincenzo Vicari, Giancarlo Carotenuto and Mario Formato - working at the Consob, Market Division, Cash and Derivatives Department - for their useful comments.

\newpage

\appendix
\begin{center}
{\huge {\bf APPENDIX}}
\end{center}

\section{Robustness analysis of k-means: outputs for other PSEs}
\begin{figure}[h]
    \centering
    \includegraphics[width=.45\linewidth]{ 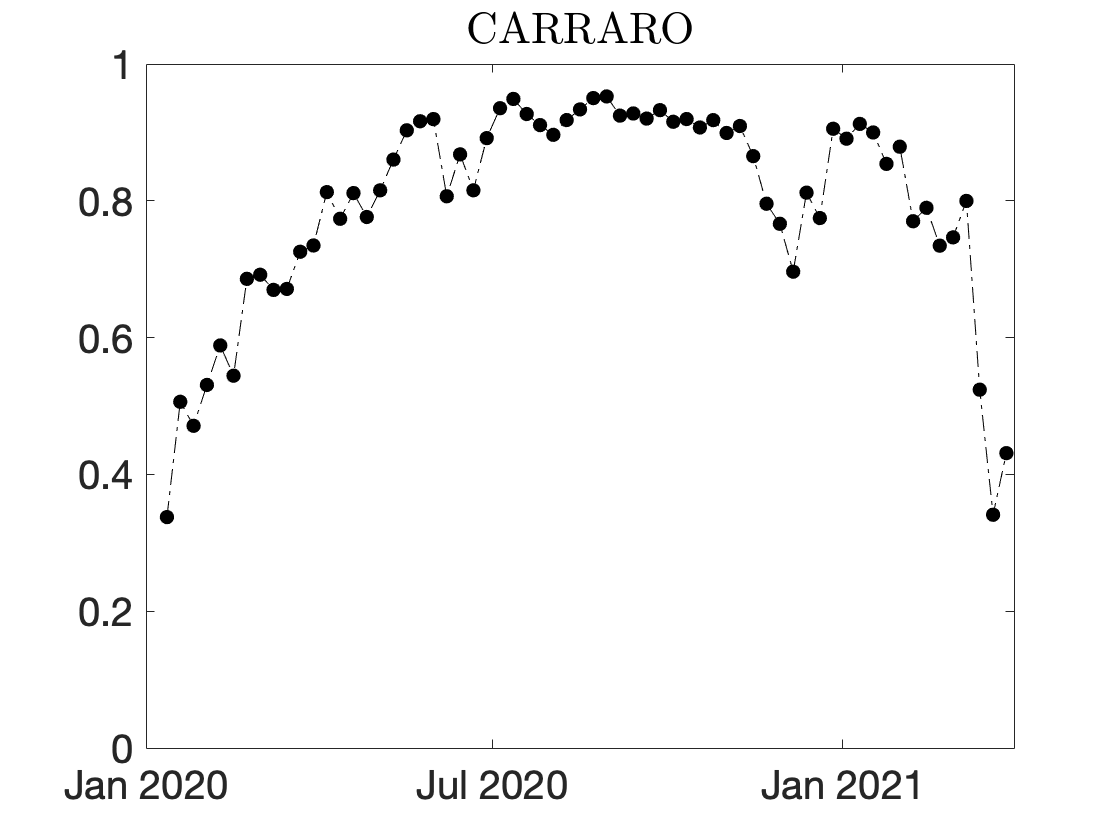}
    \includegraphics[width=.45\linewidth]{ 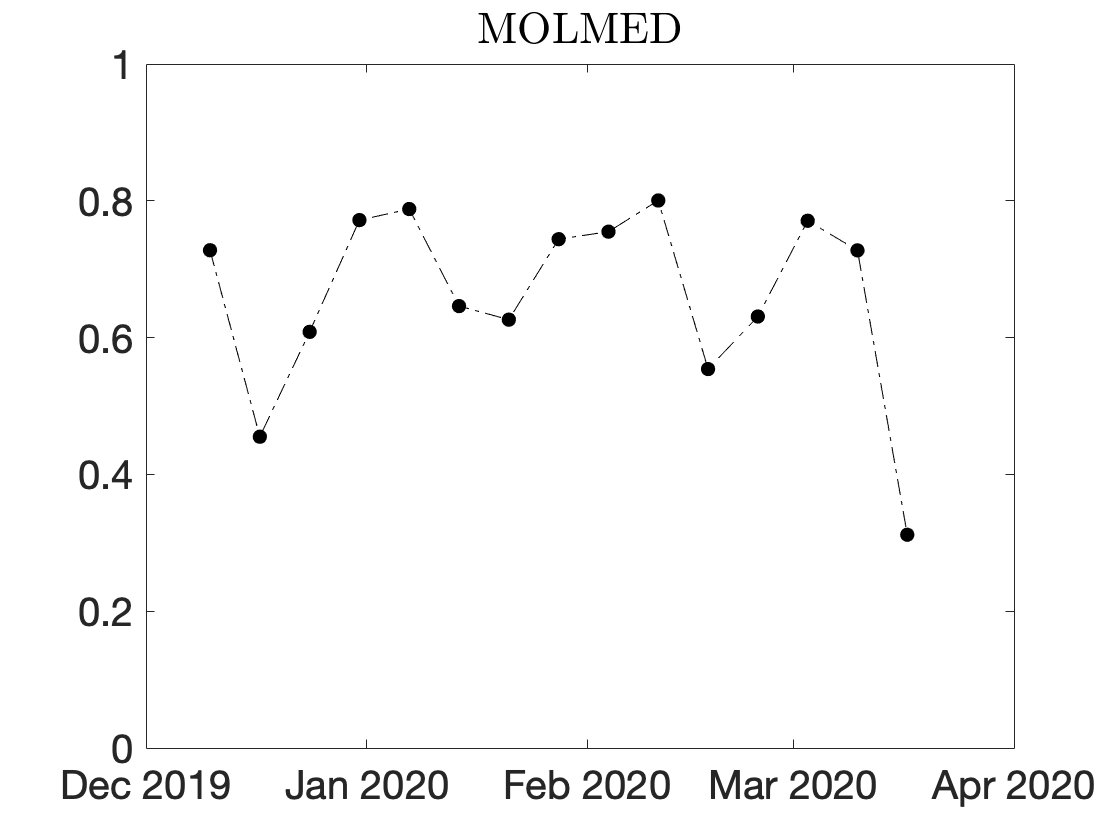}
    \includegraphics[width=.45\linewidth]{ 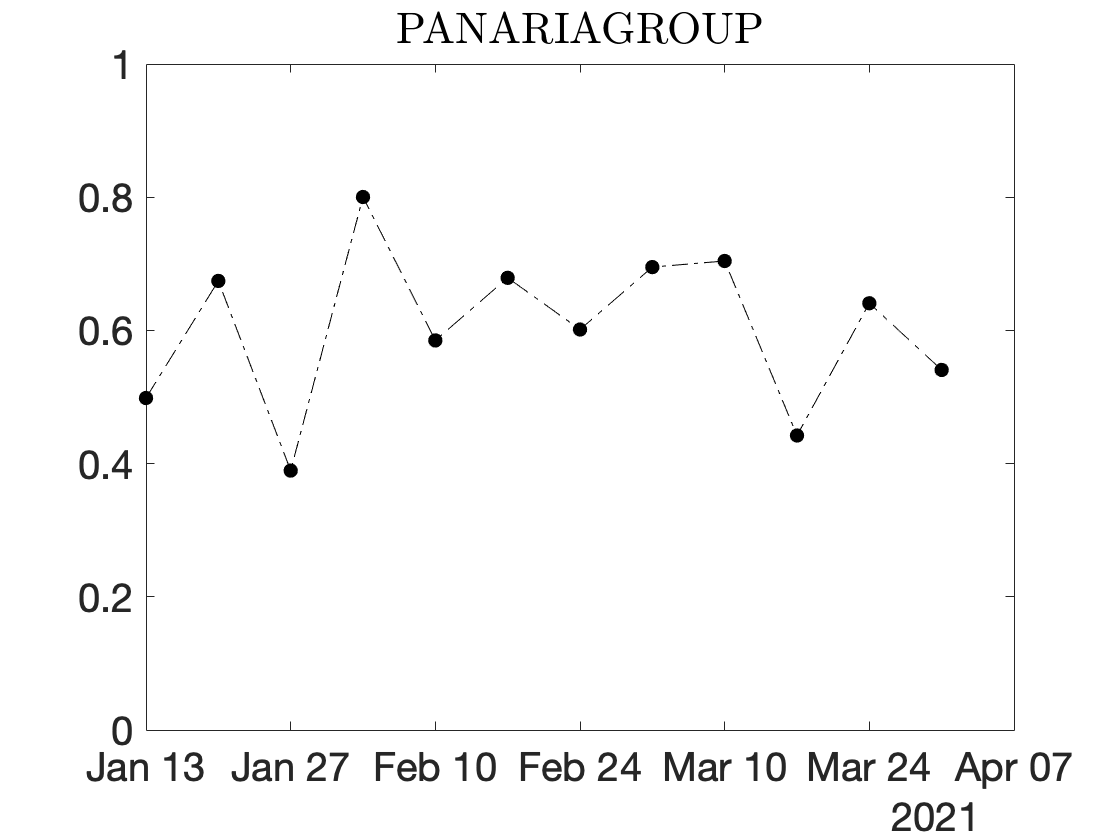}
    \includegraphics[width=.45\linewidth]{ 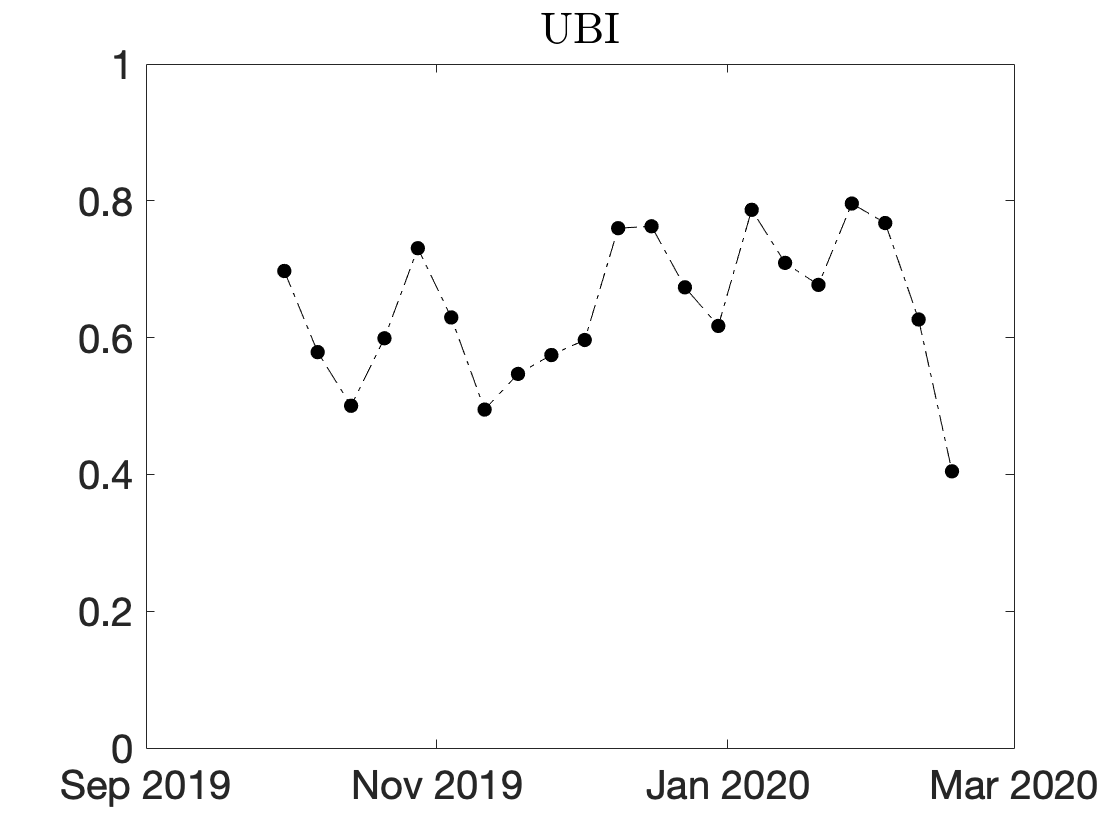}
    \caption{Jaccard similarity between two clusters in a row, by considering other takeover bid events listed in Table \ref{resPSE}}
    \label{jaccard_PSEs}
\end{figure}

\begin{figure}[h]
    \centering
    \includegraphics[width=.45\linewidth]{ 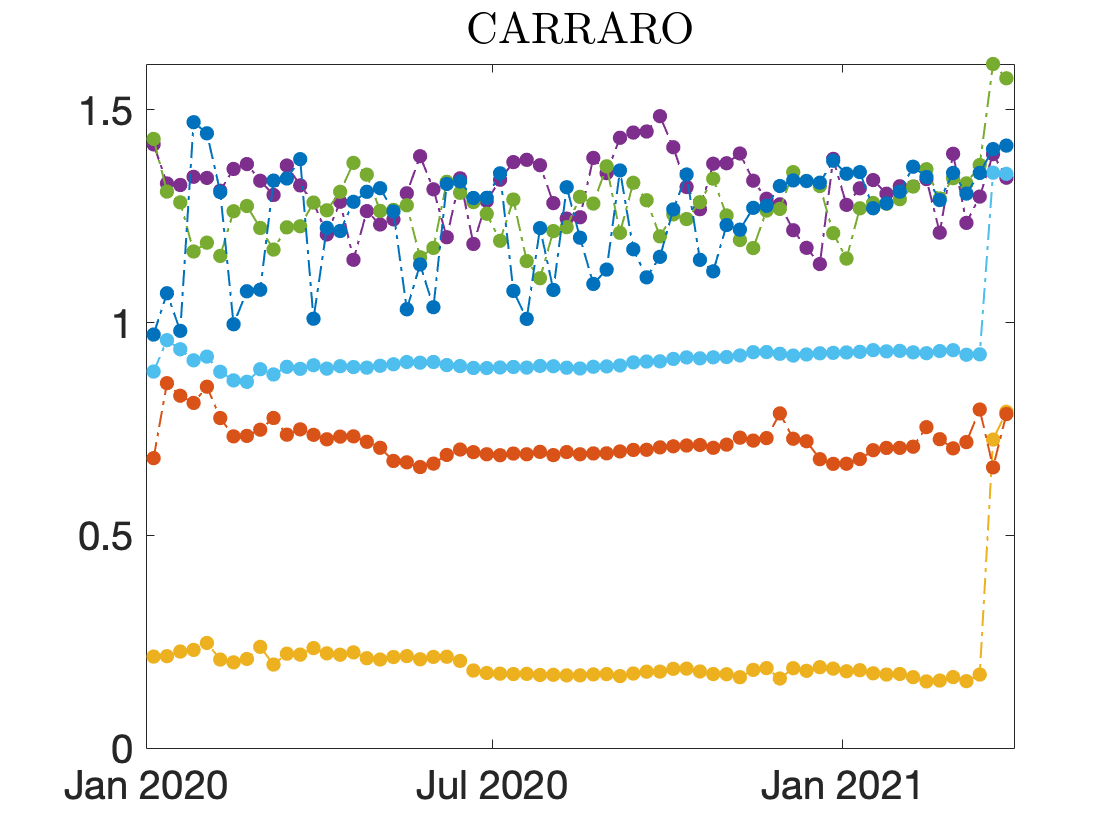}
    \includegraphics[width=.45\linewidth]{ 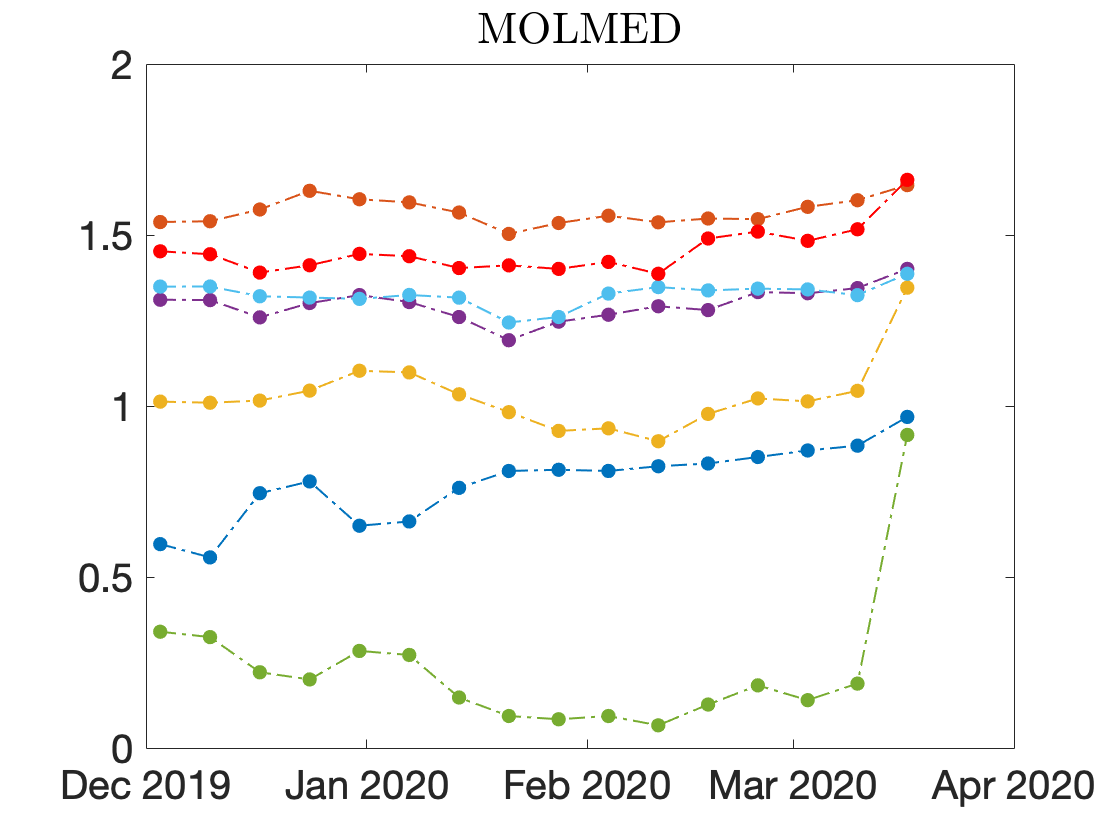}
    \includegraphics[width=.45\linewidth]{ 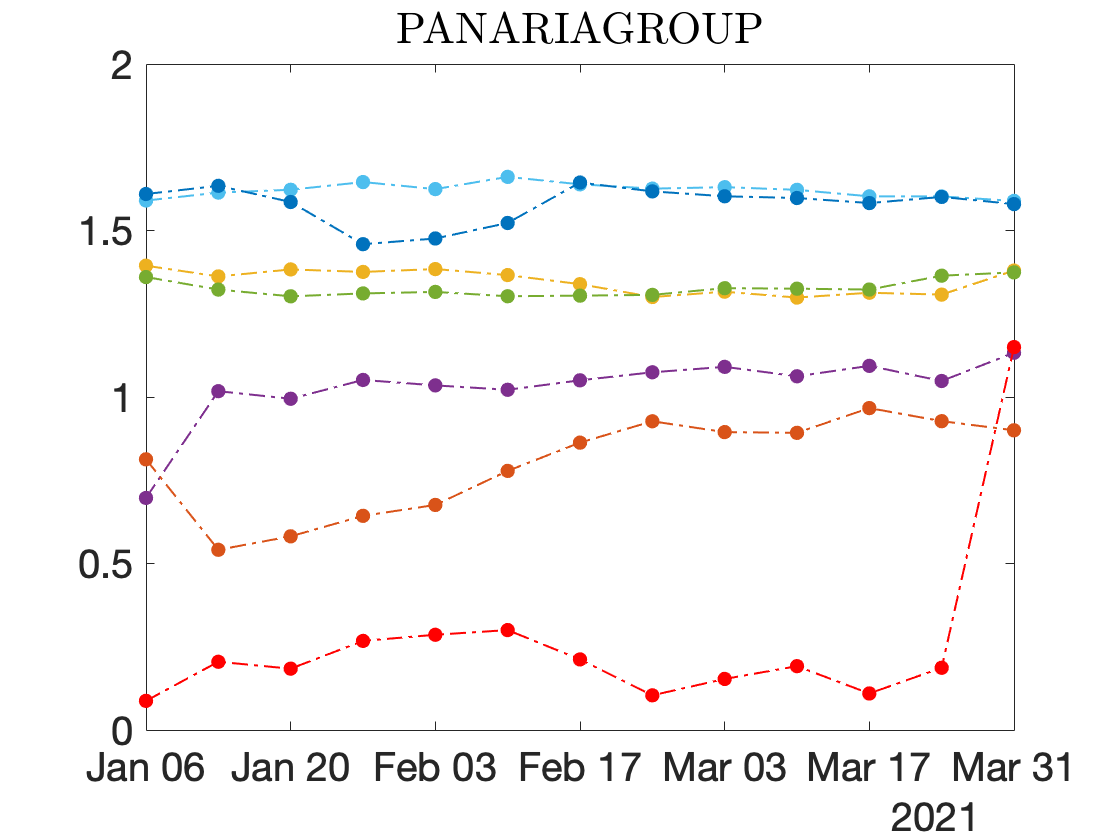}
    \includegraphics[width=.45\linewidth]{ 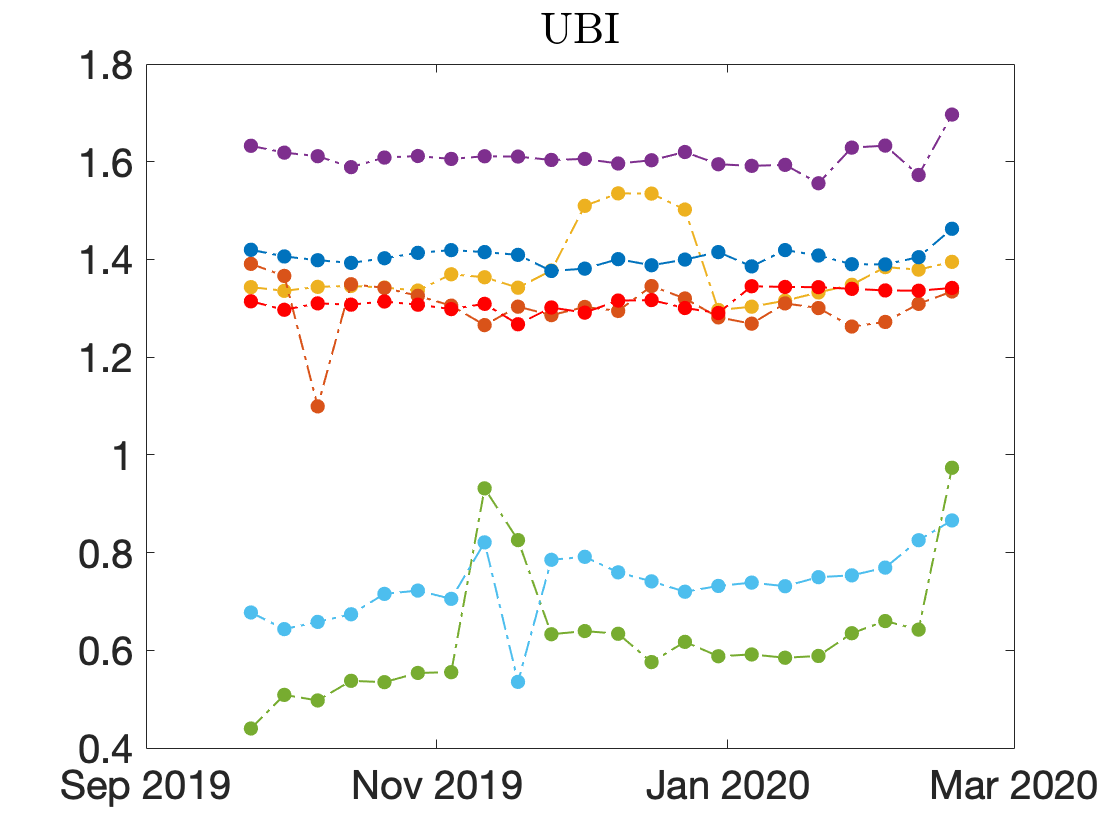}
    \caption{The evolution of the Euclidean distances of centroids from the origin, by considering other takeover bid events listed in Table \ref{resPSE}}
    \label{centroids_PSEs}
\end{figure}

The analysis of cluster stability, similarly to the one presented in Section \ref{sec_kmeans_res}, is performed also for other takeover bid events involving other stocks. The obtained results are in line with the IMA case presented above. They are summarized in Figures \ref{jaccard_PSEs} and \ref{centroids_PSEs}.

\newpage

\section{SVN: more details about the IMA case}\label{appendix_details_ima}
In table \ref{ima_traders_category_svn}, the number of different types of non-isolated traders in the Bonferroni and FDR SVN $bb$-$ss$-$bsbs$ is reported.

In tables \ref{infomap_ima_Bonferroni} - \ref{infomap_ima_FDR}, summary statistics about the most populated clusters obtained via Infomap are reported. In particular, the method illustrated in the last paragraph of subsection \ref{subsubsection_clusters} is employed to obtain the over/under-expression of traders' attributes in clusters.

\begin{table}[]
    \centering
    \small
    \begin{tabular}{c|c|c}
         Type & Bonferroni & FDR \\
         \hline
         Households & $2,287$ & $4,277$\\
         Inv. firms & $24$& $72$\\  
         Legal entities & $123$  & $324$\\
         \hline
         Total &  $2,434$ & $4,673$
    \end{tabular}
    \caption{Number of traders in the Bonferroni and FDR SVN $bb$-$ss$-$bsbs$ divided by type.}
    \label{ima_traders_category_svn}
\end{table}

\begin{table}[]
    \centering
    \small
    \begin{tabular}{c|c|c|c|c|c}
         Cluster & Traders & OI & UI& OC & UC \\
         \hline
         1 & $2098$  & H & IF, L & $ss$ & $bsbs$  \\
         2 & $65$  & &  & & $ss$ \\
         3 &  $65$ &  & & & $ss$ \\
         4 & $17$ & IF, L & H & & \\
         5 &  $16$ & L & H & & \\
         6 &$16$  & IF & H & $bsbs$ & $bb$, $ss$ \\
         7 & $9$ & L & H & & $ss$ \\
         8 & $5$ & L & H & &  \\
         9 & $8$ & L &H& & $bb$\\
         12 & $5$ & IF &H& $bsbs$ & $ss$\\
    \end{tabular}
    \caption{Summary statistics of the $10$ most populated clusters obtained by running Infomap on the Bonferroni SVN $bb$-$ss$-$bsbs$. OI/UI = over/under-expressed investor type, OC/UC = over/under-expressed co-occurrence. H = household, L = legal entity, IF = inv. firm.}
    \label{infomap_ima_Bonferroni}
\end{table}

\begin{table}[]
    \centering
    \small
    \begin{tabular}{c|c|c|c|c|c}
         Cluster & Traders & OI & UI& OC & UC \\
         \hline
         1 & $4417$ & H & IF, L & $ss$ & $bsbs$  \\
         2 & $119$ & IF, L & H &$bb$ & $ss$\\
         3 & $105$ & IF &H & $bsbs$ & $bb$, $ss$ 
    \end{tabular}
    \caption{Summary statistics of the $3$ most populated clusters obtained by running Infomap on the FDR SVN $bb$-$ss$-$bsbs$. OI/UI = over/under-expressed investor type, OC/UC = over/under-expressed co-occurrence. H = household, L = legal entity, IF = inv. firm.}
    \label{infomap_ima_FDR}
\end{table}

\begin{figure}
    \centering
    \includegraphics[width = 12cm]{ 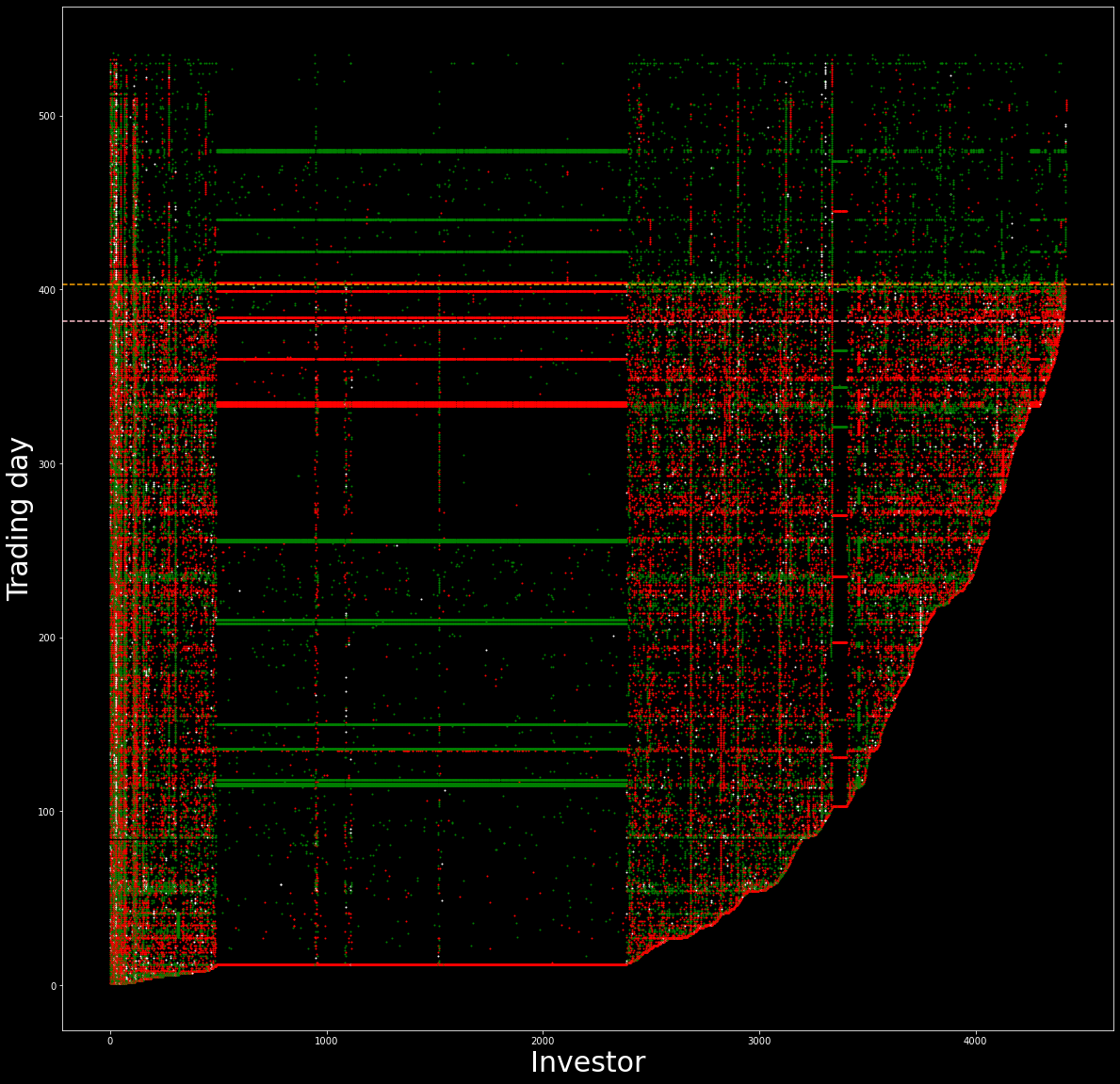}
        \caption{Graphical representation of traders' activity in the cluster $1$ obtained by running Infomap on the FDR SVN $bb$-$ss$-$bsbs$. The $x$-axis represents investors, the $y$-axis trading days. Black points correspond to no-activity, red to $b$ state, green to $s$ state and white to $bs$ state. The dotted orange horizontal line marks the PSE and the dotted pink horizontal line marks the beginning of the time window $\bar{\Delta}$.}
    \label{states1_ima_FDR}
\end{figure}
\begin{figure}
    \centering
    \includegraphics[width = 12cm]{ 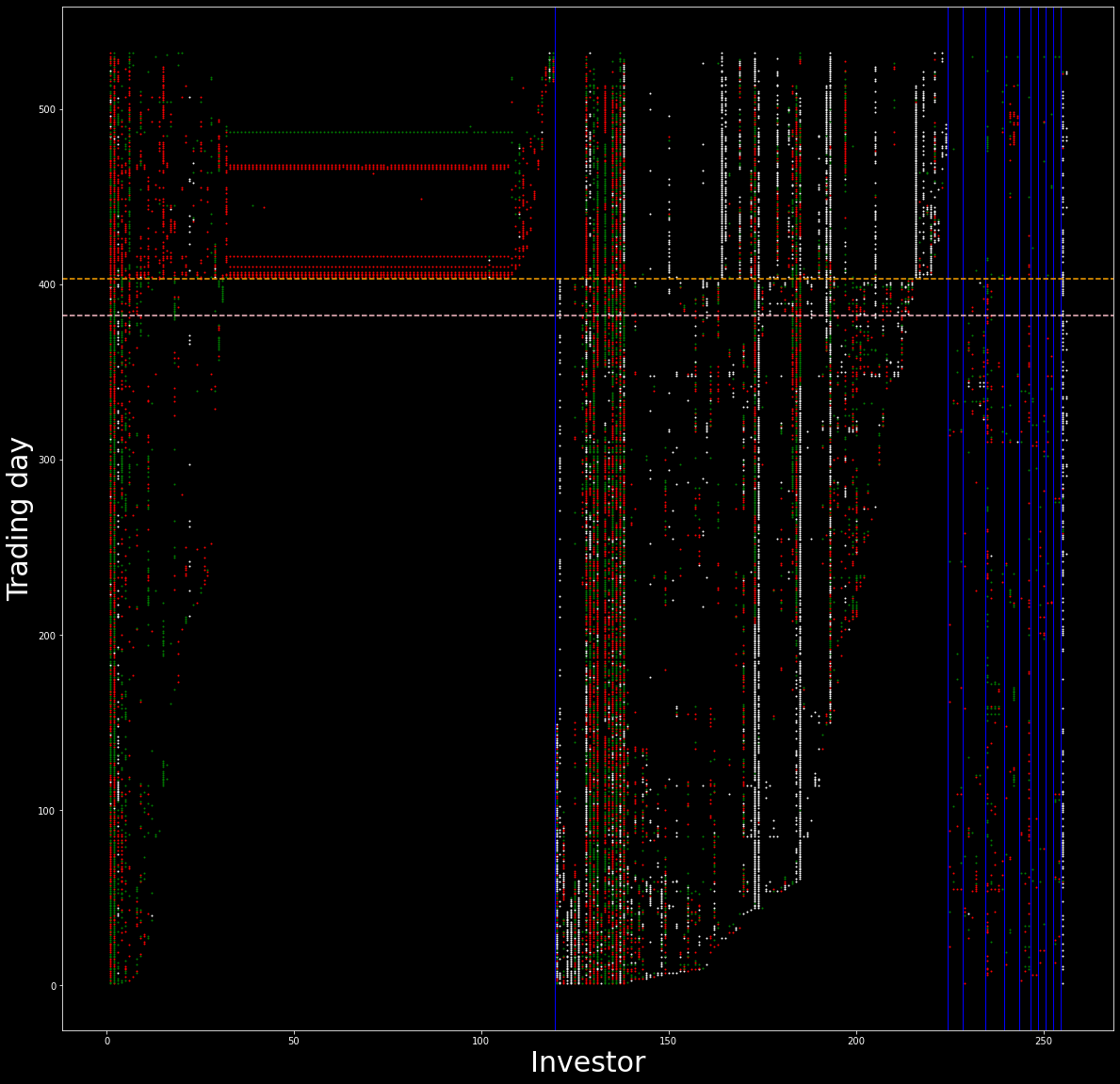}
        \caption{Graphical representation of traders' activity in the cluster $2-13$ obtained by running Infomap on the FDR SVN $bb$-$ss$-$bsbs$. The $x$-axis represents investors, the $y$-axis trading days. Black points correspond to no-activity, red to $b$ state, green to $s$ state and white to $bs$ state. Vertical light blue lines separate the clusters, the dotted orange horizontal line marks the PSE, the dotted pink horizontal line marks the beginning of the time window $\bar{\Delta}$.}
    \label{states2-end_ima_FDR}
\end{figure}

In figures \ref{states1_ima_FDR} - \ref{states2-end_ima_FDR}, plots which represent trading activity of non-isolated nodes in the FDR SVN $bb$-$ss$-$bsbs$ are displayed.

\begin{figure}
    \centering
    \includegraphics[width = 10cm]{ 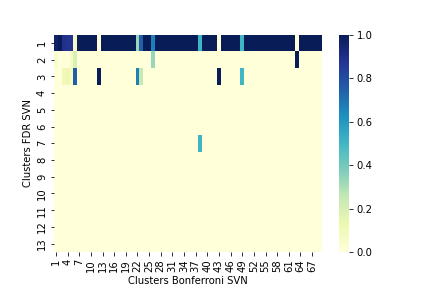}
        \caption{The element $(i,j)$ is the fraction of nodes of Bonferroni cluster $j$ that is contained in FDR cluster $i$.}
    \label{IMA_bonferroni_vs_fdr}
\end{figure}

Being the FDR correction less restrictive than Bonferroni, the SVN obtained with the latter is contained in the SVN achieved with the former. Figure \ref{IMA_bonferroni_vs_fdr} shows an heat map representing how the clusters obtained via Infomap from the Bonferroni SVN are contained in the clusters coming from the analogous FDR network. The element $(i,j)$ in the plot is the fraction of nodes of Bonferroni cluster $j$ that is contained in FDR cluster $i$. This means each column sums up to $1$ and represents how the nodes in the corresponding Bonferroni cluster are rearranged in the FDR clusters. The FDR cluster number $1$ is made up of $4,417$ elements and it turns out to contain most of the non-isolated nodes detected by the Bonferroni SVN. 

\newpage

\section{Robustness analysis of SVN: comparison with maximum entropy methods}\label{appendix_bicm}
In order to check the robustness of the SVN method for market abuse detection, the clustering is performed also with another method introduced in \cite{saracco2017}. It is an entropy-based approach which employs the Bipartite Configuration Model (BiCM) as statistical benchmark.

\subsection{Method}
As for the SVN method, the starting point is to compute the states $\{s(i,t) \ i=1,\ldots,N, \ t = 1,\ldots,T\}$ and then, to organize our data in a bipartite network where 
\begin{itemize}
    \item one layer is made up of traders: $A = \{1,\ldots,N\}$;
    \item the other layer is made up of trading days: $B = \{1,\ldots,T\}$;
    \item only links of the type $(i,t)$ $i \in A$, $t \in B$ are admitted;
    \item each link can be $b$, $s$, $bs$, depending on $s(i,t)$.
\end{itemize}

Given the bipartite network, traders similarity is computed and its statistical significance is measured by performing multiple hypothesis tests with the BiCM as benchmark.

For simplicity, let us consider our network as it just had a single type of diagonal link e.g. $bb$ and thus, let us focus on the states of type $b$. Traders similarity is defined as the number of trading days in which $i$ is in state $b$ and so does $j$:
\begin{equation*}
    N_{ij} = \sum_{t=1}^T\sigma_{it}\sigma_{jt} = \sum_{t=1}^T N_{ij}^t
\end{equation*}
where $\sigma_{it} = \mathbb{I}[(i,t) \in E]$ and $E$ is the edge set of the bipartite graph. This measure of similarity represents the number of the so-called V-motifs.

In order to quantify the statistical significance of traders' similarity, the Exponential Random Graph (ERG) class of null-models is considered. These models assign to a bipartite graph $M$ a probability 
\begin{equation*}
    P(M) = \frac{e^{-H(\theta,C(M))}}{Z(\theta)}
\end{equation*}
where $\theta$ is a vector of unknown parameters, $C(M)$ is a vector of constraints, $H(\theta,C(M))$ is the system's Hamiltonian and $Z(\theta)$ is the partition function.

In \cite{saracco2017}, several ERG models are considered; here, we choose to focus on the Bipartite Configuration Model (BiCM). The BiCM Hamiltonian imposes constraints on the degree sequences of both layers indeed, 
\begin{equation*}
    H(\theta, C(M)) = \sum_{i=1}^N \alpha_i k_{i} + \sum_{t=1}^T \beta_t h_{t}
\end{equation*}
where $k_{i}, \ i=1,\ldots,N$ and $h_{t}, \ i=t,\ldots,T$ are the degrees of traders and trading days respectively. More precisely, we have
\begin{equation*}\begin{split}
    k_{i} = \sum_{t=1}^T \sigma_{it} \\
    h_{t} = \sum_{i=1}^N \sigma_{it} \ .
\end{split}
\end{equation*}
The parameters $\alpha_{i}, \ i=1,\ldots,N$ and $\beta_{t}, \ t=1,\ldots,T$ are Lagrange multipliers which are determined by Maximum Likelihood Estimation (MLE) starting from the biadjacency matrix $M^*$ of an observed network.

The linear constraints of the system allow us to rewrite $P(M)$ in a factorized form:
\begin{equation*}
    P(M) = \prod_{i=1}^N\prod_{t=1}^T p_{it}^{\sigma_{it}} (1-p_{it})^{1 - \sigma_{it}}
\end{equation*}
where
\begin{equation*}
    p_{it} = \frac{e^{-(\alpha_i + \beta_t)}}{1 + e^{-(\alpha_i + \beta_t)}} \ .
\end{equation*}

The presence of linear constraints in the Hamiltonian amounts at treating links as independent random variables. This means $N_{ij}$ is the sum of $T$ independent Bernoulli random variables with
\begin{equation*}
    \begin{split}
        \mathbb{P}(N_{ij}^t = 1) = p_{it}p_{jt} \\
        \mathbb{P}(N_{ij}^t = 0) = 1 - p_{it}p_{jt} \ . 
    \end{split}
\end{equation*}
Thus, $N_{ij}$ is a Poisson-Binomial random variable and 
\begin{equation*}
    \mathbb{P}(N_{ij} = n) = \sum_{C \subset C_n} \Bigg[ \prod_{t \in C} p_{it}p_{jt} \prod_{t' \notin C} (1-p_{it'}p_{jt'})\Bigg]
\end{equation*}
where $C_n$ is the set of all subsets made up of $n$ integers that can be selected from $\{1,2,\ldots,T \}$.

Given this distribution, the computation of the p-value follows:
\begin{equation*}
    p(N^*_{ij}) = \mathbb{P}(N_{ij} \geq N^*_{ij}) \ 
\end{equation*}
where $N^*_{ij}$ is the value of the V-motifs in the observed network.

As for the SVN method, the link $(i,j)$ is validated whether $p(N^*_{ij})$ is lower than a statistical threshold which is corrected with the FDR procedure.

In order to implement this method, we used the Python package \textit{bicm}, which relies on the algorithm introduced in \cite{hong} to compute the Poisson-Binomial distribution. Once the validated projected network of traders is obtained, we performed the clustering with Infomap.

\subsection{Results}
As for the SVN method, first we obtain activity states and the bipartite network of traders and trading days. Given this network, V-motifs are validated as it is described in the previous paragraph. Our ultimate goal is to perform clustering on the validated projected network of traders with only diagonal links therefore, the maximum entropy method can be run on a reorganized version of the bipartite network: it is made up of three disjoint bipartite graphs, each one characterized by a different edge type ($b$, $s$ and $bs$).

So, the maximum number of edges in the projected network of traders we obtain with this approach, is $3N(3N-1)/2$. This number is greater than the corresponding one we have in the SVN method i.e. $9N(N-1)/2$ ; indeed, in that case the validated network was obtained allowing for all links and not-diagonal edges were removed secondly. However, we observe $3N(3N-1)/2$ and $9N(N-1)/2$ share their dominating terms and given $N=4,844$, the corrections in multiple tests for the two methods have a difference which is negligible.

\begin{figure}
    \centering
    \includegraphics[width = 10cm]{ 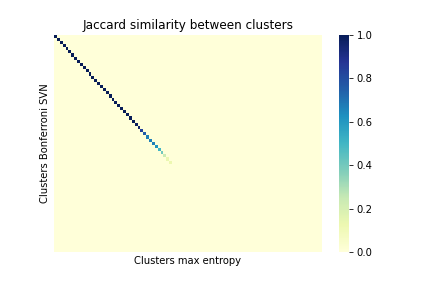}
        \caption{Heat map of the Jaccard similarity between clusters obtained with the Bonferroni SVN and the max entropy method. Clusters in the $x$ and $y$ axis are not in ordinal order: they are rearranged such that their correspondence is more evident.}
    \label{jaccard_similarity_bicmvsBonferroni}
\end{figure}

The validated projected network of traders that is obtained with the maximum entropy method has $3,279,920$ edges and $2,751$ non-isolated nodes. After running Infomap, the clusters are $93$ and analyses similar to the ones carried out in the SVN method can be done. However, we would like to focus on a comparison of the clusters obtained by the Bonferroni SVN and the maximum entropy method. In figure \ref{jaccard_similarity_bicmvsBonferroni}, an heat map representing the Jaccard similarity \cite{tibi} between couples of clusters obtained with the two procedures is reported. The line which can be identified, shows that basically, $41$ clusters have a one-to-one correspondence and among them, $31$ have values of Jaccard similarity greater than $0.8$.

\end{document}